\documentclass[amsmath,amssymb,floatfix,10pt,aps,pre,superscriptaddress,notitlepage]{revtex4-1}

\makeatletter
\renewcommand{\p@subsection}{}
\renewcommand{\p@subsubsection}{}
\makeatother

\newcommand{\bvec}[1]{{\mathbf{\string#1} }}
\newcommand{\upd}{\mathrm{d}}
\usepackage{graphicx}
\usepackage{amssymb,amsfonts,amsmath}
\usepackage{color}
\usepackage{ulem}

\newcommand{\fu}[2]{\ensuremath{\left\{\begin{array}{c}#1\\#2\end{array}\right.}}
\newcommand{\fuA}[2]{\ensuremath{\left.\begin{array}{l}#1\\#2\end{array}\right.}}

\newcommand{\pvec}{\bvec{p}}
\newcommand{\rr}{\bvec{r}}
\newcommand{\nab}{\boldsymbol{\nabla}}
\newcommand{\xxi}{\boldsymbol{\xi}}
\newcommand{\xchi}{\boldsymbol{v}}
\newcommand{\eeta}{\boldsymbol{\eta}}

\newcommand{\taua}{\tau_\text{a}}
\newcommand{\taur}{\tau_\text{R}}

\newcommand{\ttau}{\tilde{\tau}}
\newcommand{\Da}{\mathcal{D}_\text{a}}

\DeclareSymbolFont{matha}{OML}{txmi}{m}{it}
\DeclareMathSymbol{\varv}{\mathord}{matha}{118}
\newcommand{\Vext}{\varv}
\newcommand{\PsiF}{f} 
\newcommand{\sigmad}{d}
\newcommand{\di}{\mathfrak{d}}

\newcommand{\Hess}{{\nab\nab}}

\newcommand{\UCNA}{\text{(u)}}
\newcommand{\FOX}{\text{(f)}}

\begin{document}

\title{Effective equilibrium states in the colored-noise model for active matter I. 
Pairwise forces in the Fox and unified colored noise approximations}
\author{Ren\'e Wittmann}
 \affiliation{Department of Physics, University of Fribourg, CH-1700 Fribourg, Switzerland}
\author{C.\ Maggi}
\affiliation{NANOTEC-CNR, Institute of Nanotechnology, Soft and Living Matter Laboratory, Piazzale A. Moro 2, I-00185, Roma, Italy }
\author{A.\ Sharma}\affiliation{Department of Physics, University of Fribourg, CH-1700 Fribourg, Switzerland}
\affiliation{Leibniz-Institut f\"ur Polymerforschung Dresden, D-01069 Dresden, Germany}
\author{A.\ Scacchi}\affiliation{Department of Physics, University of Fribourg, CH-1700 Fribourg, Switzerland}
\author{J.\ M.\ Brader}\affiliation{Department of Physics, University of Fribourg, CH-1700 Fribourg, Switzerland}
\author{U.\ Marini Bettolo Marconi}
\affiliation{Scuola di Scienze e Tecnologie, 
Universit\`a di Camerino, Via Madonna delle Carceri, 62032, Camerino, INFN Perugia, Italy}
\date{\today}

\begin{abstract}
The equations of motion of active systems can be modeled in terms of 
Ornstein-Uhlenbeck processes (OUPs) with appropriate correlators.
For further theoretical studies, these should be approximated to yield a 
Markovian picture for the dynamics and a simplified steady-state condition.
We perform a comparative study of the Unified Colored Noise Approximation (UCNA)
and the approximation scheme by Fox recently employed within this context.
We review the approximations necessary to define
 effective interaction potentials in the low-density limit 
and study the conditions for which these  represent the behavior observed in two-body simulations 
for the OUPs model and Active Brownian particles.
The demonstrated limitations of the theory for potentials with a negative slope or curvature can be qualitatively corrected 
by a new empirical modification. 
In general, we find that in the presence of translational white noise the Fox approach is more accurate.
Finally, we  examine an alternative way to define a force-balance condition
in the limit of small activity. 
\end{abstract}

\maketitle

\tableofcontents

\section{Introduction}
Active Brownian particles (ABPs) provide a simple, minimal model system to study the 
collective behavior of active matter. Many-body Brownian dynamics simulations of these systems 
have provided considerable insight into a range of interesting nonequilibrium phenomena, 
such as the accumulation of particles at 
boundaries~\cite{Elgeti,Kantsler,Xiao,Yang,Ni} and motility-induced phase 
separation~\cite{CatesReview}. 
Much of the phenomenology of ABPs can be captured using coarse-grained, hydrodynamic 
theories~\cite{CatesReview,Cates_PNAS,fily,SpeckLoewen,gonnella2015motility},
which do not contain all information about the interparticle correlations. 
Some progress has recently been made in the linear response regime,
which allows to decouple the equations of motions of the one-body density and polarization vector~\cite{SharmaBrader}.  

Due to the inherent difficulty of dealing simultaneously with both the translational and 
orientational degrees of freedom in active systems, attempts to develop a first-principles 
theory have largely focused on a simpler, related model, in which the particle dynamics 
are represented by a set of coupled Ornstein-Uhlenbeck processes (OUPs). Within this 
model an exponentially correlated noise term, with a given correlation time, serves as 
proxy for the persistent trajectories of ABPs (connections between the two models were 
explored in Ref.~\onlinecite{faragebrader2015}). 
While the removal of orientational degrees of freedom does indeed simplify the problem, it comes 
at the cost that one has to deal with the non-Markovian dynamics of the translational coordinates. 
Fortunately, there exist various approximation methods \cite{grigolini0,grigolini,fox1,fox2,ucna1,ucna2,hasegawa} 
which enable the OUP 
model to be represented using an effective Markovian, and therefore tractable, dynamics. 
Two different approaches to doing this, (i) the Unified Colored Noise 
Approximation (UCNA) of H\"anggi {\it et al.} \cite{ucna1,ucna2}, based on adiabatic elimination on the 
level of the Langevin equations, and (ii) the Fox approximation \cite{fox1,fox2}, 
for which an approximate Fokker-Planck equation is developed,
have recently been adopted in the context of developing simple theoretical tools to describe active particles \cite{faragebrader2015,wittmannbrader2016,maggi2015sr,marconi2015,marconi2016mp,
marconi2016sr,marconi2016,sharma2016}.  

When applied to active matter, both the UCNA and Fox approximations are 
referred to as `effective equilibrium' approaches.
 The Markovian character of the dynamics implies that they obey a Fokker-Planck equation from which an effective probability distribution can in principle be obtained.  
Indeed, the possibility of mimicking the behavior of nonequilibrium ABPs using an 
equilibrium system of passive particles, interacting via effective interactions, was 
suggested by several researchers (see, e.g., Ref.~\onlinecite{Schwarz-Linek}) who observed that the phase 
separation induced by activity in systems of repulsive ABPs closely resembles that familiar 
from passive systems with an attractive interaction. 
Despite its appeal, several years were required before this observation could be turned into 
something more concrete. 
By starting from the simpler OUP model it became possible, via application of the UCNA \cite{maggi2015sr,marconi2015,marconi2016mp,
marconi2016sr,marconi2016} and Fox \cite{faragebrader2015,wittmannbrader2016,sharma2016}
approximations, to put the notion of an effective equilibrium description on a firmer footing.

In this paper, we will compare and contrast the two different approaches to effective 
equilibrium. We will highlight the main approximations involved and assess the validity 
of the effective-potential approximation (EPA), which has been employed in previous work to investigate 
activity-induced modifications of the microstructure and motility-induced phase 
separation~\cite{faragebrader2015}. This analysis both clarifies the nature of the approximations 
involved and suggests ways in which the description can be improved.

The paper is laid out as follows: 
In Sec.~\ref{sec_Theory} we first specify the model under consideration and describe 
the UCNA and Fox approaches to obtaining an effective equilibrium picture  
highlighting similarities and differences between them. 
In Sec.~\ref{sec_ueff} we describe in detail the EPA, 
where the emphasis is placed on the UCNA due to its simpler structure. 
The resulting approximate effective potentials are compared to computer simulations 
using a standard soft-repulsive and a non-convex (Gaussian core) potential.
In Sec.~\ref{sec_manyparticleTHEORY} we consider an alternative approach to obtain pairwise forces, i.e., the low-activity limit,
and make contact to the EPA. 
Finally, we conclude in Sec.~\ref{sec_conclusions}.

\section{Theory \label{sec_Theory}}

In this section, we introduce the common starting point of both the UCNA and Fox approach.
Since particles driven by Gaussian colored noise 
originally were not intended as a model for an active system,
the choice of parameters in the literature may
depend on the dimensionality and on whether contact to 
ABPs is made~\cite{faragebrader2015} or not~\cite{maggi2015sr}.
We will also clarify some notational issues.

\subsection{Colored-noise model \label{sec_CNM}}

We consider the coupled stochastic (Langevin) differential equations
\begin{equation}
\dot{\rr}_i(t) = \gamma^{-1}\bvec{F}_i(\rr_1,\ldots,\rr_N) + \xxi_i(t) + \xchi_i(t)
\label{eq_OUPs}
\end{equation}
of $N$ particles.
The motion of each particle $i\!\in\!\{1,\ldots,N\}$ at position $\rr_i(t)$ is determined by 
conservative $\bvec{F}_i$ and stochastic forces $\gamma\xxi_i$ and $\gamma\xchi_i$.
The friction coefficient $\gamma\!=\!(\beta D_\text{t})^{-1}$ 
is related to the translational Brownian diffusivity $D_\text{t}$ and
 $\beta\!=\!(k_\text{B}T)^{-1}$ is the inverse temperature.
We assume that the total interaction force $\bvec{F}_i(\rr^N)\!=\!-\nab_i\mathcal{U}(\rr^N)$ 
can be written as the gradient of a pairwise additive many-body potential
\begin{equation}
\mathcal{U}(\rr^N)=\left(\Vext(\rr_i)+\frac{1}{2}\sum_{\smash{k\neq i}}^Nu(\rr_i,\rr_k)\right)\,,
 \label{eq_Ftot}
\end{equation}
 consisting of the
one-body external fields $\Vext(\rr_i)$ and the interparticle potentials $u(\rr_i,\rr_k)\!=\!u(|\rr_i-\rr_k|)$.

The vector $\xxi_i(t)$ represents the translational Brownian diffusion by a Gaussian (white) noise of zero mean and
$\langle\xxi_i(t)\xxi_j(t')\rangle\!=\!2D_\text{t}\boldsymbol{1}\delta_{ij}\delta(t-t')$ with the unit matrix $\boldsymbol{1}$.
 Here and in the following the dyadic product of two vectors with $\di$ components results in a $\di\times\di$ matrix.
Any contraction as in a scalar product or a matrix-vector product will be explicitly indicated by a ``$\cdot$''.
 Hereafter, we shortly refer to the variable $\xxi_i(t)$ as (thermal) noise.
The OUPs $\xchi_i(t)$ defined by 
\begin{equation}
\dot{\xchi}_i(t)=-\frac{\xchi_i(t)}{\taua}+\frac{\eeta_i(t)}{\taua}
\label{eq_OUPsDEF}
\end{equation}
with $\langle\eeta_i(t)\eeta_j(t')\rangle\!=\!2D_\text{a}\boldsymbol{1}\delta_{ij}\delta(t-t')$
describe a fluctuating propulsion velocity as a non-Gaussian (colored) noise of zero mean 
and
\begin{equation}
 \langle\xchi_i(t)\xchi_j(t')\rangle=\frac{v_0^2}{\di}\boldsymbol{1}\delta_{ij}e^{-\frac{|t\!-\!t'|}{\tau_\text{a}}}
 =\frac{D_\text{a}}{\taua}\boldsymbol{1}\delta_{ij}e^{-\frac{|t\!-\!t'|}{\taua}}\,.
 \label{eq_OUPsCORR}
\end{equation}
Here we introduced the active time scale $\taua$ at which the orientation randomizes 
and the active diffusion coefficient $D_\text{a}\!=\!v_0^2\taua/\di$, where
$v_0^2\!=\!{\langle\xchi^2_i(t)\rangle}$ is the average squared self-propulsion velocity and $\di$ the spatial dimension.

The colored-noise model for active particles contains two parameters describing the magnitude and persistence of the self propulsion.
We now aim to clarify some notational differences in the literature. 
The persistence time $\taua$ can be
explicitly related to the equations of motion of run and tumble particles~\cite{cates2008} 
or ABPs~\cite{cates2013,faragebrader2015}.
The above definitions correspond to the latter case with $\taua\!=\!D^{-1}_\text{r}/(\di-1)$,
where $D_\text{r}$ is the rotational Brownian diffusion coefficient.
Another common choice~\cite{marconi2016mp} 
amounts to consider $\taur\!=\!D_\text{r}^{-1}$ 
and $D_\text{a}\!=\!v_0^2\taur/(\di-1)/\di$.
In the following, we use $\ttau\!:=\!\taua/\gamma\!\equiv\!\beta\tau\sigmad^2$, 
where $\sigmad$ is the typical diameter of a particle and the dimensionless persistence time $\tau=\taua D_\text{t}/\sigmad^2$ 
has been introduced in Refs.~\cite{faragebrader2015,wittmannbrader2016}.

Since the dimensionless active diffusivity $\Da\!:=\!D_\text{a}/D_\text{t}$ implicitly depends on the persistence time,
it constitutes the most general measure for the activity (together with the persistence length $v_0\taua$).
In order to connect to a system of ABPs~\cite{faragebrader2015,wittmannbrader2016},
it is convenient to consider instead of $\Da$ a dimensionless velocity $Pe\!=\!v_0d/D_\text{t}$,
i.e., the Pecl\'et number.
In the literature, some other definitions of a Pecl\'et number are used, which we will not consider here.
 One peculiar property of a system of active OUPs is that,
even at vanishing self-propulsion velocity $v_0\!=\!0$, or $\Da\!=\!0$, there is a contribution of $\xchi_i(t)$ to Eq.~\eqref{eq_OUPs} arising from a finite reorientation time $\taua$~\cite{sharma2016}.
One thus does not recover the equation of motion of a passive (Brownian) particle, as in the ABPs model.
In the long-time limit, however, the contribution to the dynamics becomes irrelevant and the same steady state is described as for a passive Brownian particle,
see appendix~\ref{app_SIM} for more details.
A proper passive system can only be recovered from Eq.~\eqref{eq_OUPs} only in the limit $\taua\!\rightarrow\!0$,
in the sense that the velocity correlation in Eq.~\eqref{eq_OUPsCORR} reduces to a white noise.
A Brownian system is then represented by $\Da\!=\!1$ when the thermal-noise variable $\xxi_i(t)$ is removed,
or, trivially, by setting $\Da\!=\!0$ which amounts to neglect the contribution of the OUPs.

\subsection{Effective equilibrium approach \label{sec_EEA}}

The most important step towards a theoretical study of the OUPs model
is to derive from the non-Markovian stochastic process \eqref{eq_OUPs} 
an equation of motion for the $N$-particle probability distribution $\PsiF_N(\rr^N,t)$. 
In this section, we will discuss the differences between the 
multidimensional generalizations 
of the UCNA~\cite{ucna1,ucna2} and the Fox~\cite{fox1,fox2} approaches to effective equilibrium
and expound the surprising similarities between these two approximations in 
the (current-free) steady state.

As a central quantity emerging in both cases,
we define the $\di N\times\di N$ friction tensor $\Gamma_{[N]}$ with the components
\begin{align}
\Gamma_{ij}(\rr^N)=\boldsymbol{1}\delta_{ij}-\ttau\nab_i \bvec{F}_j=
\delta_{ij}\Gamma_{ii}(\rr^N) +(1-\delta_{ij})\ttau\nab_i\nab_j u(\rr_i,\rr_j)\ \ \ \ 
\label{eq_Gamma0}
\end{align}
resulting in the Hessian of $\mathcal{U}$
and the diagonal $\di\times\di$ block
\begin{equation}
\Gamma_{ii}(\rr^N):=\boldsymbol{1}+\ttau\nab_i\nab_i\bigg(\Vext(\rr_i)+\sum_{k\neq i}^N u(\rr_i,\rr_k)\bigg)
\label{eq_Gamma1}
\end{equation}
not to be confused with $\Gamma_{[1]}(\rr_1)$ for $N\!=\!1$ particle.
In the following, we  briefly denote by $\Gamma_{ij}^{-1}$
the $ij$th block component of the inverse tensor $\Gamma_{[N]}^{-1}$.

The UCNA~\cite{maggi2015sr,marconi2015} amounts to explicitly inserting the OUPs \eqref{eq_OUPsDEF} 
into the overdamped limit of the time derivative of \eqref{eq_OUPs}, resulting in the
modified Langevin equation 
$\dot{\rr}_i(t)\!=\!\Gamma^{-1}_{ij}(\rr^N)\left(\gamma^{-1}\bvec{F}_j(\rr^N) + \xxi_j(t) + \eeta_j(t)\right)$.
It is now straight-forward to obtain for this (approximate) Markovian system driven by white noise
the Smoluchowski equation 
$\partial\PsiF_N(\rr^N,t)/\partial t=-\sum_{i=1}^N\nab_i\cdot\bvec{J}_i(\rr^N,t)$
with the probability current (the superscript $\UCNA$ denotes that the UCNA has been used)
\begin{equation}
\bvec{J}^{\UCNA}_i=\sum_kD_\text{t}\Gamma^{-1}_{ik}\cdot\bigg(\beta\bvec{F}_k\PsiF_N
-(1+\Da)\sum_j\nab_j\cdot\left(\Gamma^{-1}_{jk}\PsiF_N\right)\!\bigg)\,.
\label{eq_JUCNA}
\end{equation}
 Note that the UCNA remains valid as long as the friction tensor \eqref{eq_Gamma0}
 is positive definite.

The Fox approximation scheme applied to \eqref{eq_OUPs}, on the other hand, 
only makes use of the correlator \eqref{eq_OUPsCORR} of the OUPs,
which, in turn, may also be interpreted as the correlator of $\xchi_i(t)\!\simeq\!v_0\bvec{p}_i(t)$ 
corresponding to a coarse-grained equation of motion representing ABPs
with a constant velocity $v_0$ in the direction of their instantaneous orientation $\bvec{p}_i$
that is subject to Brownian rotational diffusion~\cite{faragebrader2015}.
This method directly yields the approximate Smoluchowski equation (superscript $\FOX$) with~\cite{sharma2016,SpeckCRIT}
\begin{equation}
\bvec{J}^{\FOX}_i=D_\text{t}\bigg(\beta\bvec{F}_i\PsiF_N-\nab_i\PsiF_N-\Da\sum_j\nab_j\cdot(\Gamma^{-1}_{ji}\PsiF_N)\bigg)\,,
\label{eq_JFOX}
\end{equation}
where the regime of validity is the same as for UCNA.
The major difference between Eq.~\eqref{eq_JUCNA} and~\eqref{eq_JFOX} 
only affects the effective description of the dynamics as a result of 
the additional factor $\Gamma^{-1}_{ik}$ arising on the level of the Langevin equation within the UCNA.
Note that in the original generalization of the Fox result~\cite{faragebrader2015} 
the tensor from Eq.~\eqref{eq_Gamma0} was incorrectly obtained as $\Gamma_{ij}\!\approx\!\delta_{ij}(1-\ttau\nab_i\cdot \bvec{F}_i)$,
which we will later identify as the \textit{(diagonal) Laplacian approximation}.
It will turn out that this (or another) approximation is necessary to obtain 
physical expressions for the effective interaction potentials.

\subsection{Two versions of the steady-state condition \label{sec_EEAss}}

In contrast to the dynamical problem, the (current-free) steady-state conditions 
\begin{equation}
\beta\bvec{F}_i P_N-\sum_j\nab_j\cdot(\mathcal{D}_{ji}P_N)=0
\label{eq_ss1}
\end{equation}
for the stationary distribution $P_N(\rr^N)$ can be cast in a coherent form, 
defining the effective diffusion tensor $\bvec{D}_{[N]}(\rr^N)=D_\text{t}\mathcal{D}_{[N]}(\rr^N)$, 
such that only the components
\begin{align}
\mathcal{D}^\UCNA_{ij}(\rr^N)&:=(1+\Da)\,\Gamma^{-1}_{ij}(\rr^N)\,, \label{eq_GammaU}\\
\mathcal{D}^\FOX_{ij}(\rr^N)&:=\boldsymbol{1}\delta_{ij}+\Da\,\Gamma^{-1}_{ij}(\rr^N)\,.\label{eq_GammaF}
\end{align}
differ between the UCNA $\UCNA$ and Fox $\FOX$ results.

Multiplying Eq.~\eqref{eq_ss1} with $\mathcal{D}^{-1}_{ik}$ and summing over repeated indices, 
the steady-state condition takes the more instructive (approximate) form~\cite{marconi2015}
\begin{align}
0=\sum_i\mathcal{D}^{-1}_{ik}\cdot\beta\bvec{F}_i P_N-\nab_kP_N-P_N\nab_k\ln|\det\mathcal{D}_{[N]}|=:\beta\bvec{F}_k^\text{eff}P_N-\nab_kP_N
\label{eq_ss2}
\end{align}
 introducing the effective force $\bvec{F}_k^\text{eff}(\rr^N)$.
 The term $\nab_k\ln|\det\mathcal{D}_{[N]}|$ is an approximation for $\sum_{ij}\mathcal{D}^{-1}_{ik}\cdot\nab_j\cdot\mathcal{D}_{ji}$,
 which becomes exact in the UCNA~\cite{marconi2015}. 
 For the Fox approach, we argue in appendix~\ref{app_DET} that this is still true in some important special cases, 
 such that Eq.~\eqref{eq_ss2} is accurate enough for our purpose.
 For high particle numbers $N$ the contribution of the off-diagonal elements to 
$\mathcal{D}_{[N]}$ becomes increasingly irrelevant~\cite{marconi2015},
which amounts to setting $\mathcal{D}_{ij}\!\rightarrow\!\delta_{ij}\mathcal{D}_{ij}$.
Assuming this \textit{diagonal form}, the determinant in Eq.~\eqref{eq_ss2} can be replaced according to
$\det\mathcal{D}_{[N]}\!\rightarrow\!\det\mathcal{D}_{kk}$
  as we have $\sum_{ij}\mathcal{D}^{-1}_{jk}\cdot\nab_j\cdot\mathcal{D}_{ij}\!\equiv\!
  \mathcal{D}^{-1}_{kk}\cdot\nab_k\cdot\mathcal{D}_{kk}\!\approx\!\nab_k\ln|\det\mathcal{D}_{kk}|$ 
  before approximating the expression in the last step (compare appendix~\ref{app_DET}).

Putting aside the dynamical behavior, described in Sec.~\ref{sec_EEA}, 
the only difference between the UCNA or Fox approximation is manifest in the definitions,
\eqref{eq_GammaU} and \eqref{eq_GammaF}, of $\mathcal{D}_{[N]}$.
Using UCNA the active diffusivity $D_\text{a}$ only appears as part of a prefactor in \eqref{eq_GammaU},
so that the friction matrix $\Gamma_{[N]}$, representing a correction due to activity,
contributes to the steady-state result even in the case $D_\text{a}\!=\!0$, that is when
$v_0\!=\!0$ and $\taua\!\neq\!0$.
Hence, the logical parameter suggested by the UCNA to tune the activity is $\taua$, 
with the passive system ($\mathcal{D}_{ij}\!=\!\boldsymbol{1}\delta_{ij}$) restored only in the limit $\taua\!\rightarrow\!0$.
 This appears to be an artifact of the pathological contribution of $\taua$ to the displacement of the OUPs, whereas
 the connection to the experimentally more relevant system of ABPs is lost.
For the latter it appears more natural to tune $v_0$ at constant $\taua$.
In the derivation of the Fox result \eqref{eq_GammaF}, on the other hand,
the explicit time evolution of the OUPs in Eq.~\eqref{eq_OUPsDEF} is irrelevant, 
suggesting a better approximate representation of ABPs~\cite{faragebrader2015}.
This reflects that we recover the (same) passive system for either
$v_0\!=\!0$ or $\taua\!=\!0$ (in the presence of noise).

Ignoring the noise contribution for $D_\text{a}\!\gg\! D_\text{t}$, 
the UCNA and Fox approximations practically describe the equivalent effective steady states.
 The major advantage of this approximation, or the UCNA result in general, is
that the inverse $\mathcal{D}_{[N]}^{-1}\!\propto\!\Gamma_{[N]}$ is pairwise additive,
even if $D_\text{t}\!\neq\!0$.
Then the effective many-body potential $\mathcal{H}_{[N]}$ 
defined as $\bvec{F}_k^\text{eff}(\rr^N)\!=\!-\nab_k\mathcal{H}_{[N]}(\bvec{r}^N)$
can be written in a closed form~\cite{maggi2015sr,marconi2015},
admitting the explicit solution $P_N(\rr^N)\propto\exp(-\beta\mathcal{H}_{[N]}(\bvec{r}^N))$
of Eq.~\eqref{eq_ss1}.
Due to the more nested form of Eq.~\eqref{eq_GammaF} 
the Fox approximation does in general not admit an analytic result.
As $\mathcal{H}_{[N]}$ is not pairwise additive in either approach,
some further approximations will become necessary to construct a predictive theory,
 which we discuss in the following sections.

\section{Effective-potential approximation (EPA) \label{sec_ueff}}
    
    Regarding the possible applications using standard methods of equilibrium liquid-state theory
a desirable strategy is to approximate $\bvec{F}_k^\text{eff}$ in Eq.~\eqref{eq_ss2} in terms of pair potentials.
 This approach allows 
to describe the phase behavior of ABPs
approximated as particles propelled by a set of coupled OUPs,
which has been discussed in detail~\cite{faragebrader2015,wittmannbrader2016}
for passive soft-repulsive and Lennard-Jones interactions in three dimensions.
However, it can be criticized that
(I.i) a system which obeys detailed balance 
is used to represent the interactions in an active system,
(I.ii) the validity criteria of the underlying theory might be violated so that further approximations are required and
(I.iii) higher-order particle interactions are neglected, which are believed to be important for
the phase separation in an active system.
In the following, we define the effective pair interaction and motivate different approximations,
which we compare to computer simulations of two ABPs and
two particles propelled by OUPs.
It is our objective to comment on the aforementioned points
and illustrate the qualitative differences between the Fox and UCNA.
For the sake of simplicity, we will restrict the presentation of technical aspects to the UCNA results.

\subsection{Calculation of effective potentials \label{sec_ueff1}}

To identify an effective pair potential $u^\text{eff}(r)$, 
we consider $N\!=\!2$ interacting particles, i.e., the low-density limit of Eq.~\eqref{eq_ss2}.
Ignoring the external forces for now by setting $\Vext(\rr)\!\equiv\!0$,
it is easy to verify that
\begin{align}
\nab_1\beta u^\text{eff}(r)
=\left(\mathcal{D}^{-1}_{11}-\mathcal{D}^{-1}_{21}\right)\cdot\nab_1\beta u(r)+\nab_1\ln|\det\mathcal{D}_{[2]}|
\label{eq_ueff}
\end{align}
and analog equation for $\nab_2u^\text{eff}(r)$, where we used $\nab_2u(r)\!=\!-\nab_1u(r)$ and $r\!=\!|\rr_1-\rr_2|$.
Keeping in mind that we seek to employ this effective potential to approximately represent the interaction of many particles,
it appears undesirable that an equal statistical weight is put to both the diagonal $\mathcal{D}^{-1}_{11}$ 
and the off-diagonal components $\mathcal{D}^{-1}_{21}$ of the diffusion tensor.
As an alternative we propose the effective potential 
\begin{align}
\nab_k\beta u^\text{eff}_\text{diag}(r)=\mathcal{D}^{-1}_{kk}\cdot\nab_k\beta u(r)+\nab_k\ln|\det\mathcal{D}_{kk}|\,,
\label{eq_ueffDIAG}
\end{align}
with $k\in\{1,2\}$, obtained for a diagonal form of $\mathcal{D}_{[2]}$ with $\Vext(\rr)\equiv0$.
For completeness we find in the one-particle limit a quite similar formula
\begin{align}
\nab\beta \Vext^\text{eff}(\rr)=\mathcal{D}^{-1}_{[1]}\cdot\nab\beta\Vext(\rr)+\nab\ln|\det\mathcal{D}_{[1]}|
\label{eq_Veff}
\end{align}
for the effective external field $\Vext^\text{eff}(\rr)$, 
since for $N\!=\!1$ we have $\mathcal{D}_{[1]}\!=\!\mathcal{D}_{11}$.
Note that a quite different expression for an effective external potential can be derived 
starting from the equations of motion for ABPs~\cite{wittmannbrader2016,pototsky2012}.

Integration of the above equalities yields the desired formulas for the effective potentials 
depending only on the bare potential $u(r)$ or $\Vext(\rr)$ and the activity parameters $D_\text{a}$ and $\taua$~\cite{faragebrader2015}.
Alternatively, we could have directly defined~\cite{marconi2015,marconi2016mp} $\Vext^\text{eff}(\rr)\!:=\!\mathcal{H}_{[1]}(\bvec{r})$
and $u^\text{eff}(r)\!:=\!\mathcal{H}_{[2]}(\bvec{r}_1,\bvec{r}_2)$ from the many-body potential $\mathcal{H}_{[N]}(\bvec{r})$
identified in the solution 
of \eqref{eq_ss1}, which is, however, inconvenient when the Fox approach is used.
Assuming a bare potential $u(r)$ obeying $\lim_{r\rightarrow\infty}u(r)\!=\!0$, the integrated form of~\eqref{eq_ueff} reads
\begin{align}
\beta u^\text{eff}(r)&=\beta\frac{ u(r)+\ttau( \partial_ru(r))^2}{1+\Da}
-\ln\!\left|E_1^{(\di-1)}(\tau,r)\,E_2(\tau,r)\right|,
\label{eq_ueffUCNA}
\end{align}
where $\partial_r\!=\!\partial/\partial r$ and
\begin{align}
 E_n(\tau,r):=1+2\ttau r^{n-2}\partial^n_ru(r)\,,\ n\!\in\!\{1,2\}
 \label{eq_EVs}
\end{align}
are the Eigenvalues of $\Gamma_{[2]}$.
We can further identify $\nab_1 u(r)$ in the first term of Eq.~\eqref{eq_ueff} as the Eigenvector 
of $\mathcal{D}^{-1}_{11}-\mathcal{D}^{-1}_{21}$ corresponding to the Eigenvalue $E_2/(1+\Da)$.
Note that in \eqref{eq_ueffUCNA} we could equally introduce an effective energy scale $\beta_\text{eff}\!=\!\beta/(1+\Da)$
to absorb the factor $(1+\Da)$~\cite{marconi2015,marconi2016mp}.
We refrain to do so as this interpretation would not be consistent with the 
the way $\Da$ enters within the Fox approach.

\subsection{Limitations and possible corrections \label{sec_ueff2}}

Studying Eq.~\eqref{eq_ueffUCNA} 
more carefully, we notice that
the effective potentials do not always behave in a physical way.
This is because, in violation of the validity condition of both  the UCNA and the Fox approximation,
the diffusion tensor $\mathcal{D}_{[2]}$ is not positive definite for a large number of relevant potentials. 
In general, we easily see that the logarithm will diverge whenever one of the Eigenvalues $E_n(\tau,r)$ vanishes. 
Given a positive and convex bare potential $u(r)\!>\!0$, the Eigenvalue $E_2$ is strictly positive,
which also means that the effective attraction solely arises from the term including the logaritm. 
However, as we have $\partial_ru(r)\!<\!0$ in this case, the Eigenvalue $E_1$ will vanish at a certain value of $r$
and we require a further approximation to remedy the unphysical behavior of $u^\text{eff}(r)$ in $\di\!>\!1$ dimensions.
At a highly non-convex or negative region of the bare potential,
the same problem occurs for $E_2$.
 Interestingly, if we only require knowledge of an effective potential on a finite interval
where the eigenvalues are positive, its overall unphysical behavior is irrelevant~\cite{sharma2016}.

 First note that there is a broader range of admissible bare potentials when the \textit{diagonal approximation},
 Eq.~\eqref{eq_ueffDIAG}, of the effective pair potential is used, 
or if we are interested in the one-body external field, Eq.~\eqref{eq_Veff}.
 This can be understood from the explicit formula for $u^\text{eff}_\text{diag}(r)$, 
 which we obtain from Eq.~\eqref{eq_ueffUCNA} 
by rescaling all terms proportional to $\tau$ with a factor $1/2$.
In the following, we propose different ways to generally
rid the effective potential of possible artifacts of vanishing Eigenvalues in the last term of Eq.~\eqref{eq_ueff}.
A correction of the first term is not necessary and also has no noticeable effect.

Let us first assume that $u(r)\!>\!0$ is convex, i.e., it represents a soft-repulsive interaction.
Then a sufficient criterion (due to the presence of the term $\boldsymbol{1}\delta_{ij}$ in Eq.~\eqref{eq_Gamma0}, 
some other potentials are allowed that are only slightly negative and slightly non-convex)
for the matrix $\Gamma_{[N]}$ to have strictly positive Eigenvalues
would be that it depends on an elliptic differential operator rather than $\nab_i\nab_j$.
Therefore, a convenient approximation is to redefine Eq.~\eqref{eq_Gamma0} by
an elliptical operator, the simplest example of which is the Laplacian $\boldsymbol{\Delta}\!=\!\nab\cdot\nab$.
Upon substituting
\begin{align}
 \nab_i\nab_j\rightarrow\boldsymbol{1}\nab_i\cdot\nab_j
\label{eq_laplacian}
\end{align}
the effective potential becomes
\begin{align}
 \beta u^\text{eff}_{\boldsymbol{\Delta}}(r)=\beta\frac{ u(r)+\ttau( \partial_ru(r))^2-2(\di-1)\int_r^\infty\upd s\,\ttau\frac{( \partial_su(s))^2}{s}}{1+\Da} -\ln\left(1+2\ttau\partial^2_ru(r)+2(\di-1)\ttau \frac{\partial_ru(r)}{r}\right)
\label{eq_ueffUCNAlap}
\end{align}
where the additional term compared to \eqref{eq_ueffUCNA} cannot be integrated in general.
This \textit{Laplacian approximation} has been successfully employed (together with the Fox and diagonal approximation)
in explicit calculations~\cite{faragebrader2015,wittmannbrader2016}.
In $\di\!=\!1$ dimensions both differential operators reduce to the second derivative and 
Eq.~\eqref{eq_ueffUCNAlap} is equal to Eq.~\eqref{eq_ueffUCNA},
which provides a good account of active particles interacting with a soft-repulsive potential~\cite{marconi2016mp}.

An alternative way is to empirically rectify 
the explicit formula for $u^\text{eff}(r)$ in Eq.~\eqref{eq_ueffUCNA}.
Most intuitively, one can expand the argument of the logarithm up to the first order in $\tau$.
In fact, this \textit{small-$\tau$ approximation} is quite similar to 
the Laplacian approximation \eqref{eq_ueffUCNAlap}
(and completely equivalent in one dimension),
but we do not recover the additional term involving the integral.
Performing the small-$\tau$ approximation of the full expression~\eqref{eq_ueffUCNA} appears too crude,
as an expansion of the logarithm  does not converge for $\ttau\sum_{l>1}\boldsymbol{\Delta} u(\rr,\rr_l)>2$.
The resulting effective potential will thus become
 totally uncontrolled for short separations of highly-repulsive particles.

A more elaborate correction that may be applied also to highly non-convex potentials
 is the \textit{inverse-$\tau$ approximation}, 
an empirical strategy maintaining the leading order in $\tau$ while not disregarding higher-order terms.
This is achieved by substituting in Eq.~\eqref{eq_ueffUCNA} $E_n(\tau,r)\!\rightarrow\!E^\text{(i)}_n(\tau,r)\!>\!0$, where
\begin{align}
 E^\text{(i)}_n(\tau,r):=\fu{1/(2-E_n(\tau,r))}{E_n(\tau,r)}\fuA{\mbox{if $E_n(\tau,r)<1$}}{\mbox{otherwise} }.
\label{eq_invtau}
\end{align}
The major advantage of this approximation is that it yields quite similar results to the full potential
whenever the validity condition is only slightly violated
 and the effective potential does not diverge if the bare potential is finite.
 The empirical motivation behind this correction is that $E_n$ constitutes the two leading terms of the ``resummed'' Taylor series of $E^\text{(i)}_n$ in the case $E_n\!<\!1$.
 As described in appendix~\ref{app_DET}, the most convenient implementation of the inverse-$\tau$ approximation for the Fox result is to identify the expressions for $E_n(\tau,r)$ in $\det\mathcal{D}^\FOX_{[2]}$ and use Eq.~\eqref{eq_invtau}.

\begin{figure}[t]{
\includegraphics[width=0.235\textwidth] {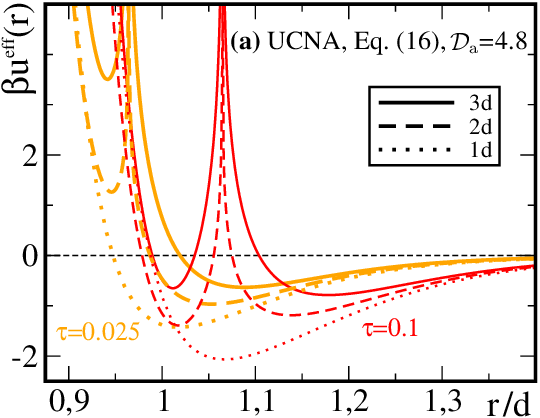} \hfill
\includegraphics[width=0.235\textwidth] {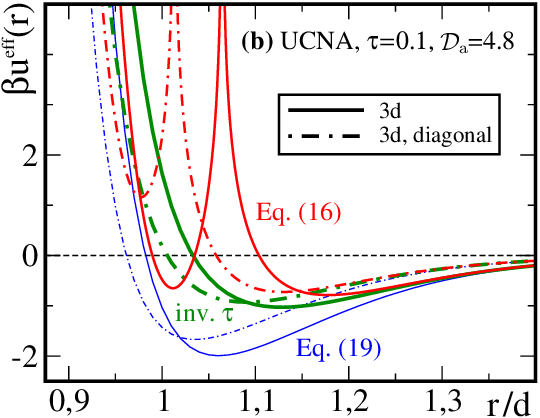} \hfill
\includegraphics[width=0.235\textwidth] {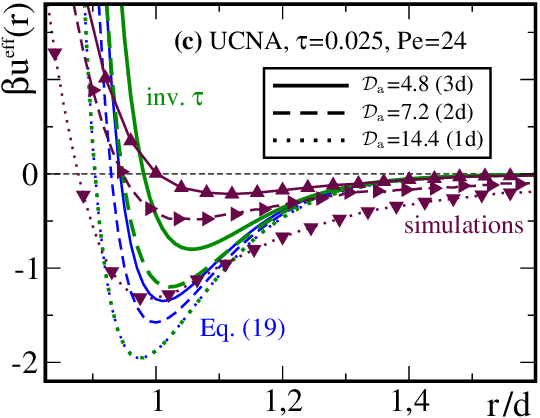} \hfill
\includegraphics[width=0.235\textwidth] {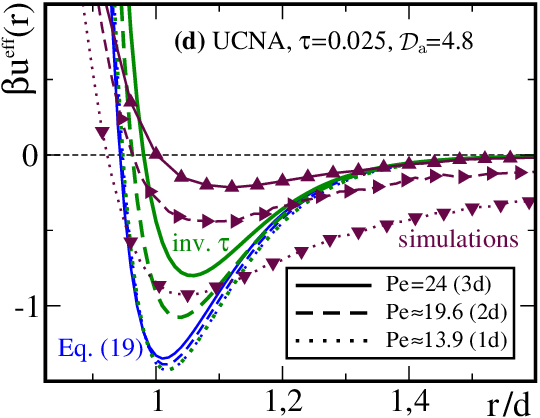}
}

\caption{Effective potentials in the presence of thermal noise for a soft-repulsive $\beta u(r)\!=\!(r/d)^{-12}$ from the UCNA in $\di\!=\!1$ (dotted lines), 
$\di\!=\!2$ (dashed lines) and $\di\!=\!3$ (solid lines) dimensions.
\textbf{(a)} Full result, Eq.~\eqref{eq_ueffUCNA}, for the active diffusivity $\Da\!=\!4.8$ 
and persistence times $\tau\!=\!0.025$ (thick, brighter lines) or $\tau\!=\!0.1$ (thin, darker lines).
\textbf{(b)} Comparison to the diagonal form (dot-dashed lines) in $\di\!=\!3$ for $\Da\!=\!4.8$ and $\tau\!=\!0.1$.
The thick lines correspond to the inverse-$\tau$ approximation, Eq.~\eqref{eq_invtau} of Eq.~\eqref{eq_ueffUCNA},
and the thin lines to the Laplacian approximation, Eq.~\eqref{eq_ueffUCNAlap}.
\textbf{(c)} Approximate results compared to simulations of active OUPS (lines with triangles) 
at $\tau\!=\!0.025$ and a constant Pecl\'et number $Pe\!=\!\sqrt{\di\Da/\tau}\!=\!24$, 
as $\Da$ increases with decreasing $\di$.
\textbf{(d)} Approximate results compared to simulations of active OUPS at $\Da\!=\!4.8$ and $\tau\!=\!0.025$.
\label{fig_UCNA}}
\end{figure}

\subsection{Comparison to computer simulations of two active particles \label{sec_SIM}}

In Sec.~\ref{sec_ueff1} we introduced different strategies 
to define a suitable effective interaction potential 
in the effective-equilibrium approximation for the colored-noise model.
Now we illustrate under which conditions an approximate treatment according to Sec.~\ref{sec_ueff2} becomes necessary 
and compare the theoretical results to computer simulations.
The easiest way to determine an effective potential numerically
is to set up a two-particle simulation, measure the radial distribution function $g(r)$ and calculate 
$\beta u^\text{eff}_{\text{sim}}(r)\!=\!-\ln g(r)$.
By doing so, we make the same approximation (I.iii) as in theory
to ignore the many-particle character of the interaction.
However, the simulations for ABPs and OUPs, detailed in appendix~\ref{app_SIM},
take into account the orientation dependence and the non-Markovian character of the dynamics, respectively.

\subsubsection{The role of approximations, dimensionality and thermal noise }

We first discuss some general observations in the UCNA for a soft-repulsive system with the bare potential $\beta u(r)\!=\!(r/d)^{-12}$.
The behavior of the Fox result is qualitatively similar.
As expected, the full expression for the effective potential in Eq.~\eqref{eq_ueffUCNA} is impractical 
as it diverges at a certain distance $r_\text{div}$,
determined by the condition $r_\text{div}\!=\!(24\tau)^{-1/12}d$,
which is when the first Eigenvalue $E_1$ within the logarithm vanishes, whereas $E_2$ is always positive.
As suggested by Fig.~\ref{fig_UCNA}a, this behavior is most problematic at larger values of $\tau$,
where $u^\text{eff}(r_\text{div})$ should be rather negative, as it is the case in $\di\!=\!1$ dimensions.
We further see in Fig.~\ref{fig_UCNA}a that this effect becomes
more severe with increasing dimension.
Both the inverse-$\tau$ and Laplacian approximations
successfully cure this unphysical divergence, which we see in Fig.~\ref{fig_UCNA}b.
As employing the diagonal form $u^\text{eff}_\text{diag}(r)$ of the effective potential 
simply amounts to a rescaling of $\tau$, we observe in Fig.~\ref{fig_UCNA}b
that it results in a smaller effective diameter of the repulsive part but a flatter potential well.
Accordingly, $r_\text{div}$ becomes smaller.

Since the definition of the active diffusivity $\Da$ depends on the dimension $\di$,
not all parameters $\tau$, $\Da$ and $Pe$ can be kept constant upon varying the dimensionality.
For a constant reorientation time $\tau$ and propulsion velocity $Pe$
the effective attraction in Fig.~\ref{fig_UCNA}c is stronger in lower spatial dimensions
for both approximations considered, which coincide with the full expression in $\di\!=\!1$.
This behavior appears sensible, as two particles have less possibilities to avoid each other upon collision,
and is in qualitative agreement with computer simulations of active OUPs.
 Moreover, we understand that motility-induced phase separation 
is harder to observe in higher dimensions \cite{stenhammar2014}.
Keeping $\Da$ constant instead of $Pe$ (which then decreases with decreasing dimension)
the same trend is observed in Fig.~\ref{fig_UCNA}d for the inverse-$\tau$ approximation and computer simulations,
whereas the result for the Laplacian approximation barely changes with dimensionality.
Comparing the qualitative behavior in Figs.~\ref{fig_UCNA}c and d, we recognize in all spatial dimensions
that the numerical effective potential is of longer range
than the theoretical predictions in any approximation.
This observation confirms the criterion discussed in Ref.~\onlinecite{ucna2}
that the UCNA is expected to become less accurate for larger separations
 where a typical length scale of the active motion,
closely related to the effective diffusion tensor, Eq.~\eqref{eq_GammaU},
exceeds the spatial scale over which the force field varies.

\begin{figure}[t]{

\includegraphics[width=0.235\textwidth] {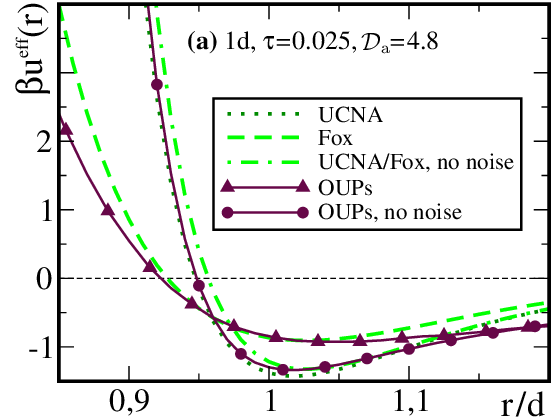} \hfill
\includegraphics[width=0.235\textwidth] {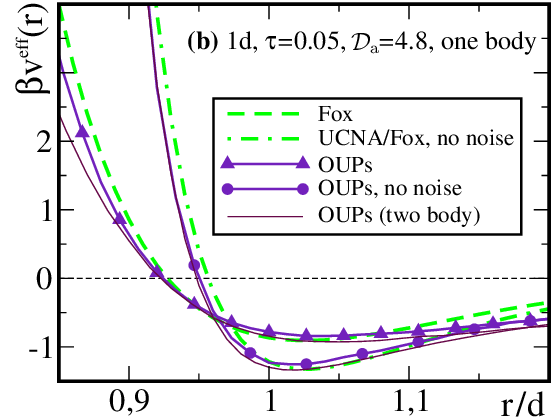} \hfill
\includegraphics[width=0.235\textwidth] {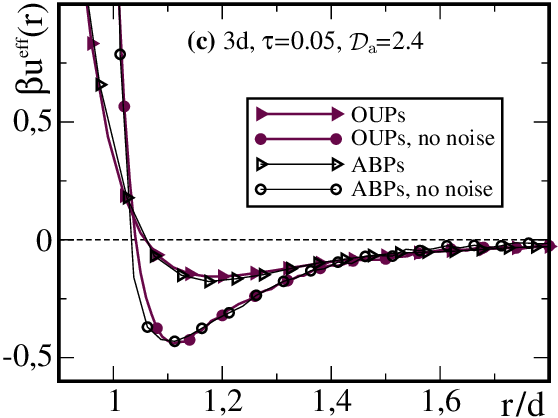} \hfill
\includegraphics[width=0.235\textwidth] {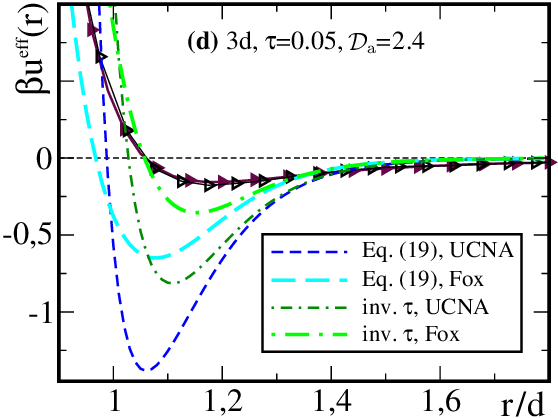}}

\caption{ Effective potentials in a soft-repulsive system with and without thermal noise.
\textbf{(a)} Comparison to simulations of active OUPS in $\di\!=\!1$ with $\Da\!=\!4.8$ and $\tau\!=\!0.025$, as in Fig.~\ref{fig_UCNA}d,
including the Fox approximation (dashed lines)
and simulations without thermal noise (labeled with dots).
In the latter case, UCNA and Fox are equivalent (dot-dashed line).
 \textbf{(b)} As Fig.~\ref{fig_noise}a, but for a single active particle in an external potential $\beta \Vext(r)\!=\!(r/d)^{-12}$ with $\tau\!=\!0.05$
chosen such that the theory predicts the same curves as for two particles with $\tau\!=\!0.025$ 
(the two-body simulation results for $\tau\!=\!0.025$ from Fig.~\ref{fig_noise}a are shown as thin solid lines for comparison).
 \textbf{(c)} Simulations of two ABPs (empty symbols) and OUPs for $\tau\!=\!0.05$ and $\Da\!=\!2.4$ 
 in $\di\!=\!3$ dimensions. \textbf{(d)} As Fig.~\ref{fig_noise}c (with noise) compared to the Laplacian approximation, Eq.~\eqref{eq_ueffUCNAlap}, and the
inverse-$\tau$ approximation as labeled. 
\label{fig_noise}}
\end{figure}

As the involved approximations become cruder in higher spatial dimensions,
 the quantitative agreement with the simulation results in Figs.~\ref{fig_UCNA}c and d becomes worse.
For $\di\!=\!1$, a remarkable agreement between the UCNA and simulation results for
the radial distribution of two particles has been reported in Ref.~\onlinecite{maggi2015sr},
where $\xxi_i(t)$ in Eq.~\eqref{eq_OUPs} was set to zero.
Doing so also in our simulations, we observe in Fig.~\ref{fig_noise}a that the effective potential deepens
and its repulsive barrier becomes steeper.
This curve is in excellent agreement with the theoretical result for zero noise,
obtained by both the UCNA and Fox approach upon dropping the first term 
in Eq.~\eqref{eq_GammaU} and Eq.~\eqref{eq_GammaF}, respectively. 
The full UCNA result is only slightly different in the repulsive regime.
Intriguingly, we also recognize in Fig.~\ref{fig_noise}a that the simulation data including the noise term 
are excellently represented by the Fox approach.
 We can understand these observations by recapitulating the idea behind the two approximations.
The UCNA amounts to manipulate Eq.~\eqref{eq_OUPs} by calculating the second derivative of $\rr_i(t)$
in order to eliminate the variable $\xchi_i(t)$ in Eq.~\eqref{eq_OUPsDEF}.
The original discussion of the accuracy of this approximation does not account for 
the second stochastic variable $\xxi_i(t)$.
In contrast, the Fox approach is only dedicated to determine the approximate contribution of the colored noise $\xchi_i(t)$ 
to the effective probability current in Eq.~\eqref{eq_JFOX}, which is independent of other terms in Eq.~\eqref{eq_OUPs}.
Therefore, the Fox theory has a broader range of applicability and
should be accurate in both the presence and the absence of thermal noise.

Finally, we note that the excellent agreement between theory and computer simulations in one dimension
implies that the diagonal approximation is not justified when it comes to describing a two-body system.
It is, however, interesting to consider a single particle in an external field of the same form as the interparticle potential considered above.
In agreement with the theoretical prediction, the computer simulations in Fig.~\ref{fig_noise}b 
show nice agreement between the two-body system and a one-body system with the double value $\tau\!=\!0.05$ of the persistence time.
We further observe that the theoretical result for one body is even closer to the simulation data than for two bodies.

\subsubsection{Soft-repulsive Brownian system in three dimensions }

\begin{figure*}[t]{
 \hfill 
 
\includegraphics[height=0.18\textwidth] {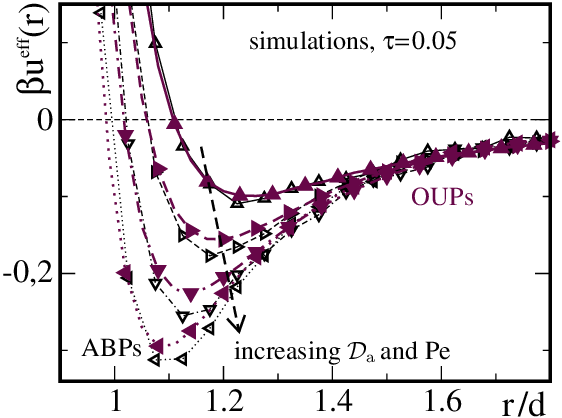} \hfill
\includegraphics[height=0.18\textwidth] {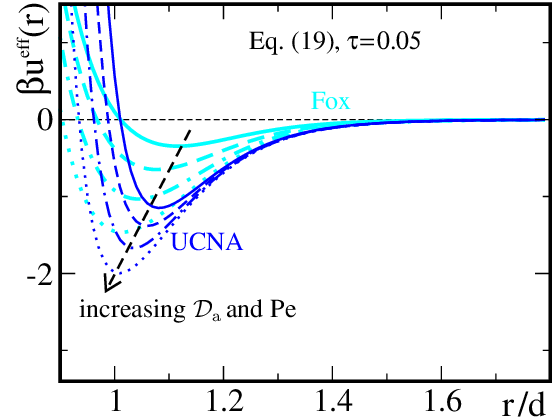}\hfill
\includegraphics[height=0.18\textwidth] {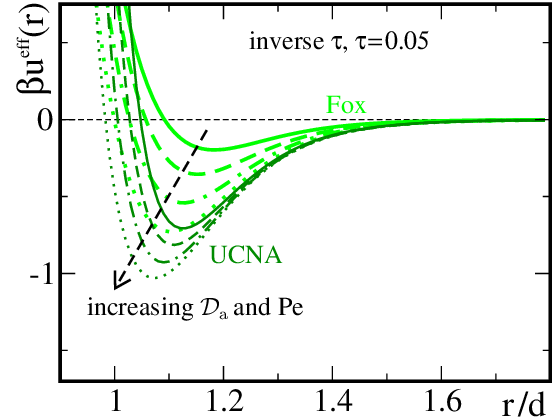}\hfill
\includegraphics[height=0.18\textwidth] {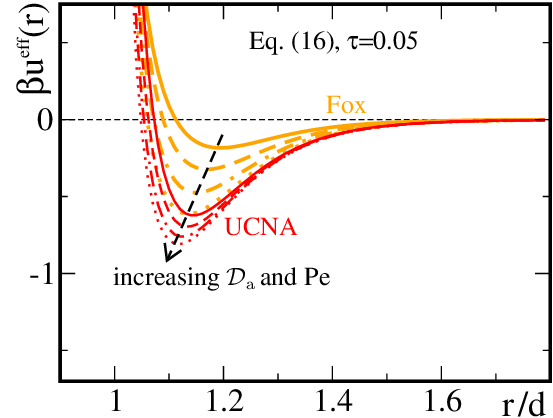}

\vspace*{.1cm}

\includegraphics[height=0.18\textwidth] {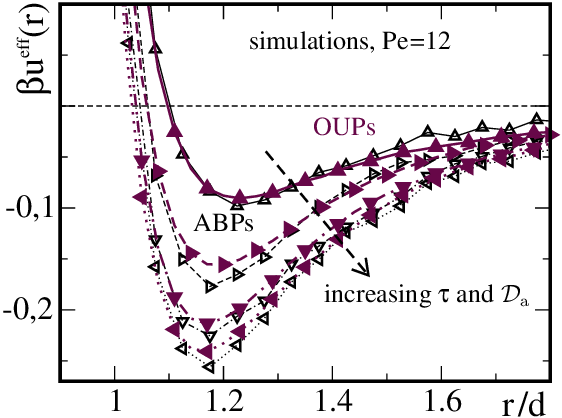} \hfill
\includegraphics[height=0.18\textwidth] {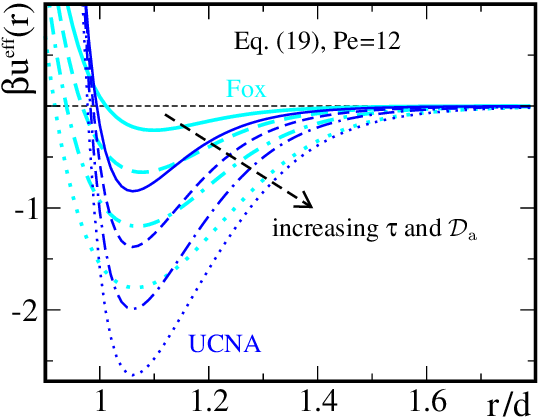} \hfill
\includegraphics[height=0.18\textwidth] {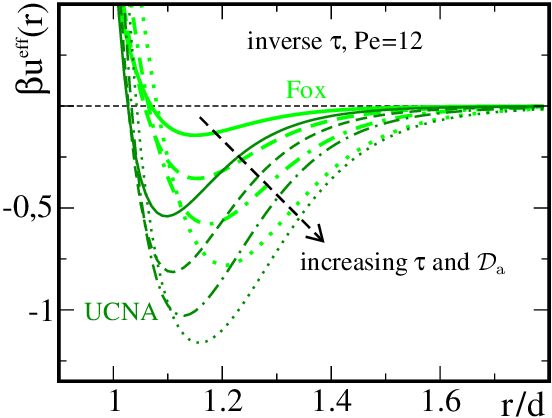} \hfill
\includegraphics[height=0.18\textwidth] {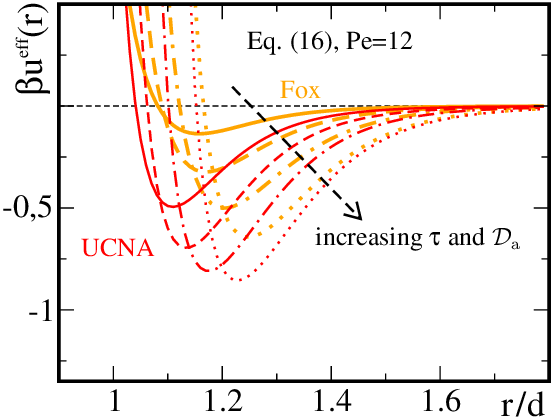}

\vspace*{.1cm}

\includegraphics[height=0.18\textwidth] {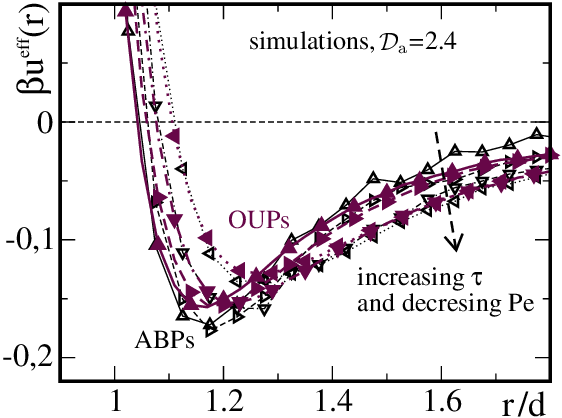} \hfill
\includegraphics[height=0.18\textwidth] {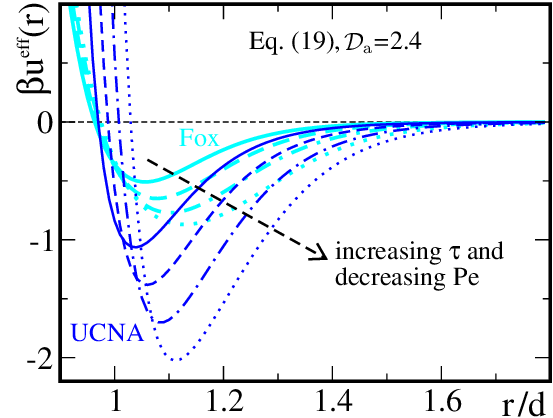} \hfill 
\includegraphics[height=0.18\textwidth] {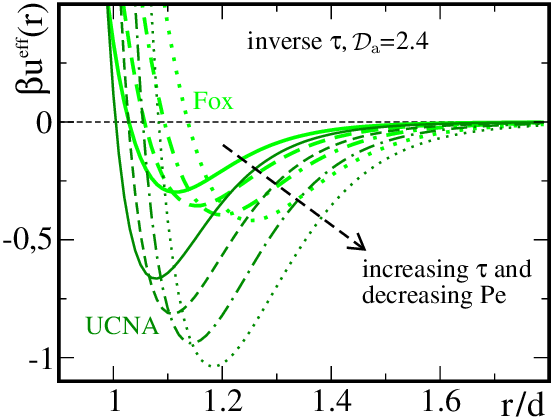}  \hfill
\includegraphics[height=0.18\textwidth] {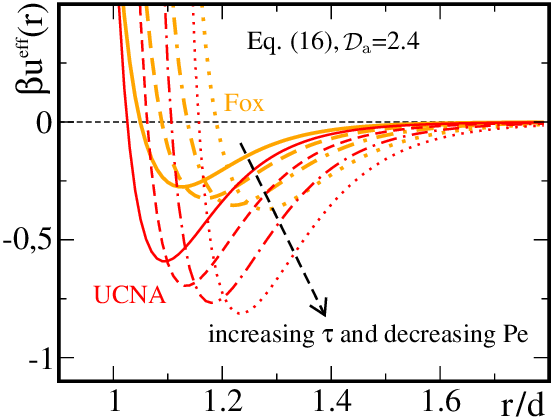}  }
\caption{
Effective potentials in three dimensions for soft-repulsive spheres, $\beta u(r)\!=\!(r/d)^{-12}$,
obtained within the Fox (thick, brighter lines) and UCNA (thin, darker lines)
and by numerical simulations of ABPs (empty triangles) and active OUPs (filled triangles).
Columns from left to right: simulations; Laplacian approximation, Eq.~\eqref{eq_ueffUCNAlap};
inverse-$\tau$ approximation, Eq.~\eqref{eq_ueffUCNA} with Eq.~\eqref{eq_invtau}; Eq.~\eqref{eq_ueffUCNA} only for $r\!>\!r_\text{div}$ (see Fig.~\ref{fig_UCNA}a for the full result).
Rows from top to bottom: increasing the active diffusivity $\Da$ at constant persistence time $\tau\!=\!0.05$;
increasing $\Da$ and $\tau$ at constant Pecl\'et number $Pe\!=\!\sqrt{3\Da/\tau}\!=\!12$;
increasing $\tau$ at constant $\Da\!=\!2.4$.
The sequence from the solid to the dotted lines corresponds to increasing the respective parameter(s)
$\Da$ from $\Da\!=\!1.2$ or $\tau$ from $\tau\!=\!0.025$ by a factor of two in each step 
(dashed lines are always for the same set of parameters).
\label{fig_3d}}
\end{figure*}

 We also performed computer simulations of ABPs (described in appendix~\ref{app_SIM}) for $\di\!=\!3$.
 For a finite active diffusivity, Fig.~\ref{fig_noise}c reveals that the numerical effective potentials for the two considered models 
 with and without noise are
nearly identical over the full range of separations.
 As for $\di\!=\!1$, the effective potential of active OUPs (and ABPs)
in the absence of thermal noise has a deeper well and a larger repulsive diameter.
Quantitatively, this difference is much more pronounced in three dimensions.
In the following, we restrict ourselves to systems with thermal noise
 and compare in Figs.~\ref{fig_noise}d and~\ref{fig_3d} to the predictions of the theory.
As in Figs.~\ref{fig_noise}c and d, our simulations of ABPs and OUPs,
shown in the first column of Fig.~\ref{fig_3d}, are in nice agreement for all sets of considered parameters.
This is quite surprising since, on the many-particle level, ABPs and OUPs have different steady states \cite{solonEPJST,szamel2014}.
On the basis of our data for the simplistic two-body system we could rather conclude that OUPs subjected to thermal noise are an excellent model for ABPs at moderate activity~\cite{faragebrader2015,wittmannbrader2016}.

 We see in Fig.~\ref{fig_noise}d that the depth of the attractive well of all theoretical versions of the effective potential in $\di\!=\!3$ dimensions
is significantly overestimated when compared to the simulations of both ABPs and OUPs.
The inverse-$\tau$ approximation appears to provide the best guess 
of the point at which the effective potential changes its sign.
 To facilitate the further qualitative comparison we chose the $y$ axes in Fig.~\ref{fig_3d} according
to the deviation from the simulations (first column), i.e., by a factor of 10 for the Laplacian approximation (second column)
and 5 for the inverse-$\tau$ approximation (third column).
The chosen approximations exhibit a behavior similar to 
the full theoretical results of Eq.~\eqref{eq_ueffUCNA} in the physical region for $r\!>\!r_\text{div}$, 
shown in the last column.
The differences arising from using the exact effective force in Eq.~\eqref{eq_ss2} are discussed in appendix~\ref{app_DET}.

For the considered soft-repulsive bare potential,
 we observe in Fig.~\ref{fig_3d} some notable quantitative differences between the UCNA and Fox results, 
 even at relatively high $\Da$.
 The effective diameter of the repulsive part is generally smaller than in the UCNA, 
 whereas the overall attraction is weaker in the Fox approach.
This becomes most apparent in the Laplacian approximation.
  The first row of Fig.~\ref{fig_3d} contains the effective potentials evolving for a constant persistence time $\tau$
  when the active diffusivity $\Da$ (or the Pecl\'et number $Pe$) is increased.
  All approaches accordingly predict an increased effective attraction 
  and the minimum of the potential is shifted to smaller separations~\cite{faragebrader2015,wittmannbrader2016}.
The Fox results exhibit a stronger variation with $\Da$, which is also more consistent with the numerical data.
Similarly, the effective potentials in the second row deepen with increasing $\tau$ at constant $Pe$,
where the location of the minimum is nearly unaffected.
The most interesting behavior is observed in the third column at constant $\Da$.
Again, all approaches agree that the minimum is shifted to larger separations with increasing $\tau$,
but the effective attraction predicted by the simulations is nearly constant,
as simultaneously the magnitude of the self-propulsion is decreased.
This observation is not consistent with the UCNA results.

Based on the presented simple comparison,
our conclusion is that the best choice for the theoretical effective potential is the Fox approach in the
inverse-$\tau$ approximation.
Upon further increasing the activity (not shown) the quantitative discrepancy of the Laplacian approximation 
becomes even more pronounced.
Regarding Fig.~\ref{fig_UCNA}b one might get the impression that the additional assumption of the diagonal form of the effective potential
also results in a slightly better (quantitative) agreement with the simulations.
However, we stress that it is not clear in how far the effective pair potentials
can accurately describe the many-body situation,
as there are no higher-order interactions present in a two-body simulation.

\subsubsection{Difficulties for non-convex potentials }

The most compelling argument in favor of the inverse-$\tau$ approximation arises from considering non-convex bare potentials,
a case in which the Laplacian approximation becomes useless above a certain value of $\tau$.
 This issue has already been discussed for a Lennard-Jones potential~\cite{SpeckCRIT}
and we will consider here the potential $\beta u(r)\!=\!\exp(-(r/d)^2)$ of an active Gaussian-core fluid. 
Although this model has not received much attention in theories for ABPs, 
it is quite appealing from a theoretical perspective.
Most prominently, this bare potential is a known exceptional case in which 
a simple mean-field theory is particularly accurate~\cite{GaussianCore1,GaussianCore2},
which might, in a way, also hold for the effective potential of the active system.

\begin{figure}[t]{
\includegraphics[width=0.235\textwidth] {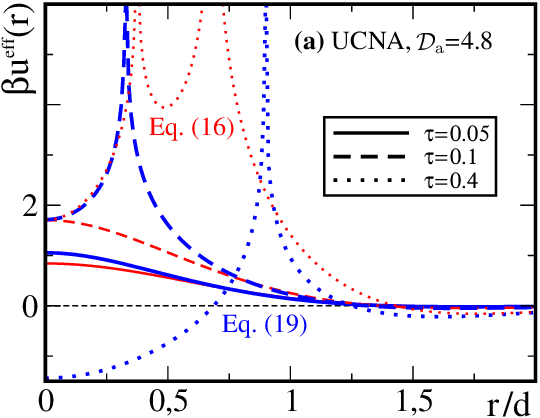} \hfill
\includegraphics[width=0.235\textwidth] {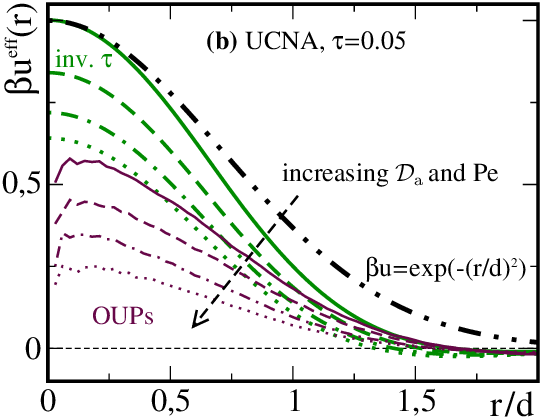} \hfill
\includegraphics[width=0.235\textwidth] {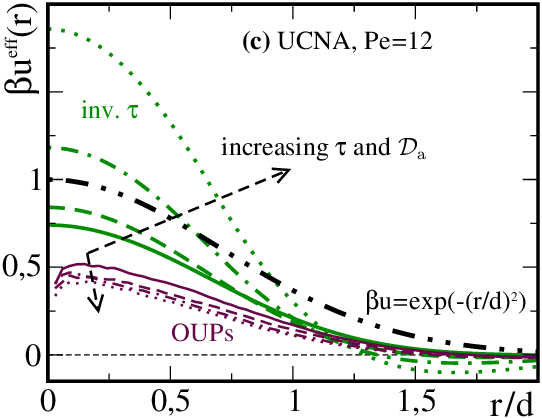} \hfill
\includegraphics[width=0.235\textwidth] {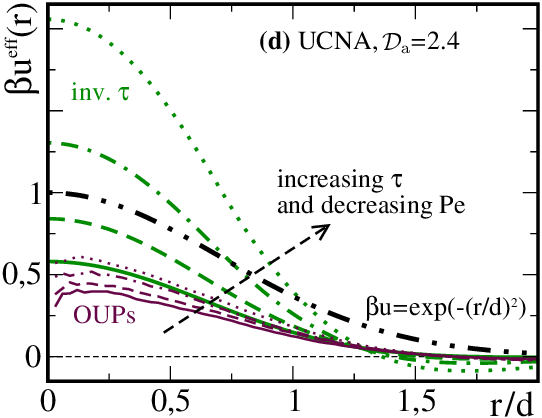}}

\caption{
Effective potentials from the UCNA in three dimensions for Gaussian-core particles, $\beta u(r)\!=\!\exp(-(r/d)^2)$. 
\textbf{(a)} Comparison of the divergences in the full result from Eq.~\eqref{eq_ueffUCNA} and the Laplacian approximation, Eq.~\eqref{eq_ueffUCNAlap}, for different persistence times $\tau$.
\textbf{(b)}-\textbf{(d)} Inverse-$\tau$ approximation
and computer simulations for active OUPs
for the same parameters as in rows 1-3 of Fig.~\ref{fig_3d}, respectively.
 The bare potential is shown as the thick dot-dot-dashed line.
\label{fig_GC}}
\end{figure}

The effective potential of an active Gaussian-core fluid is discussed in Fig.~\ref{fig_GC} within the UCNA.
As the absolute value of both the curvature and slope of this model potential is bounded, 
the Fox results (not shown) are quite similar, even for the moderate values of $\Da$ considered here.
When the persistence time $\tau$ is sufficiently small, 
the effective potential, Eq.~\eqref{eq_ueffUCNA}, does not diverge 
   and the different approximations behave in a quite similar way.
Interestingly, we observe in Fig.~\ref{fig_GC}a 
that the divergence of the Laplacian approximation sets in at an even smaller value of $\tau\!\geq\!1/12$ than for the full potential.
The latter diverges at two points, each related to one of the two Eigenvalues, if $\tau\!\geq\!1/4$.

In Fig.~\ref{fig_GC}b, c and d we discuss the only suitable form of the effective potential, i.e., 
the inverse-$\tau$ approximation.
Intriguingly, the predicted behavior depends on in which way, 
i.e., by means of which parameter, the activity is modified.
Increasing the active diffusivity (or the Pecl\'et number) at a constant value of $\tau$ 
results in a less repulsive core.
Upon increasing $\tau$, however, the height of the maximum increases and 
an attractive well develops at larger separations.
 Quite counterintuitively, we observe that in this case the effective interaction becomes more repulsive than in the passive case,
also for the inverse-$\tau$ approximation.
Our computer simulations (also carried out for values of $\tau$ much larger than shown in Fig.~\ref{fig_GC}) 
confirm that this is an artifact of the theory, 
related to the negative curvature of the bare potential.
At constant $\tau$ 
the evolution of the theoretical results agrees qualitatively with the simulations.
The simulation data are, however, not very sensitive to changes in the persistence time.
At constant $\Da$ the theory predicts the correct trend upon increasing $\tau$,
whereas this is not the case at constant Pecl\'et number.

 To argue about the validity of the EPA,
we consider two classes of bare interactions.
Firstly, soft-repulsive and convex potentials lead to quite accurate results in one dimension,
but require an empirical correction in higher dimensions.
Secondly, the understanding of the behavior 
of particles interacting with a bare potential which has a negative curvature 
remains one of the most urgent open problems in our theoretical framework.
 At the moment, the only way to obtain a workable theory in this case is by employing the inverse-$\tau$ correction introduced in Eq.~\eqref{eq_invtau}.
Further numerical and theoretical analysis will be needed to fully clarify this issue.
Finally, we note that potentials with attractive parts do not \textit{a priori} constitute a problem for the theory,
but usually have regions in which they are non convex.
For a discussion of such problems see also Refs.~\onlinecite{SpeckCRIT,sharma2016}.

\section{Low-activity approximation
\label{sec_manyparticleTHEORY}}

A second strategy to simplify the steady-state condition
is to perform an expansion in the activity parameter $\tau$.
At linear order, the effective diffusion tensor $\mathcal{D}_{[N]}(\rr^N)$
becomes pairwise additive.
In this low-activity approximation, a YBG-like hierarchy can be obtained
by successively integrating Eq.~\eqref{eq_ss1} over $N-n$ coordinates~\cite{marconi2015},
which allows defining a mechanical pressure and interfacial tension~\cite{marconi2016}.
Moreover, for the active system evolving according to Eqs.~\eqref{eq_OUPs} and~\eqref{eq_OUPsCORR} it has
been demonstrated that there exists a regime for small values of $\tau$,
where the principle of detailed balance is respected~\cite{fodor2016}.
This suggests that, at leading order in this parameter, the approximations resulting in Eq.~\eqref{eq_ss1}
are perfectly justified.

Knowing, however, that Eq.~\eqref{eq_ss1} contains the same information as Eq.~\eqref{eq_ss2},
 which depends logarithmically on the parameter $\tau$,
 we should clarify whether
 (II.i) the low-activity expansion converges,
 (II.ii) it is sufficient to only consider the leading order
 and (II.iii) one can obtain similar results when employing the EPA.
  To do so, we demonstrate how the first member ($n\!=\!1$) of the YBG hierarchy can be
 rederived from Eq.~\eqref{eq_ss2} and discuss the consequences of approximating $\bvec{F}_k^\text{eff}$ in terms of pair interactions.
Again, we only discuss the UCNA, where, without any further approximation, the inverse diffusion tensor
is found to be pairwise additive.

\subsection{Alternative derivation of a local force-balance condition}

By integrating a multidimensional vector equation (label $i$) over $N-1$ coordinates we understand 
multiplying each side by $\delta(\rr-\rr_i)$, followed by summation over all particles $i$ and integation over all $N$ spatial coordinates.
We further define the average $\langle \bvec{X}\rangle\!:=\!\langle\!\langle\sum_{i=1}^N\delta(\rr-\rr_i)\bvec{X}_i\rangle\!\rangle/\rho(\rr)$ of a vector $\bvec{X}_i(\rr^N)$,
where $\langle\!\langle\,\cdot\,\rangle\!\rangle$ denotes the full canonical ensemble average
 and $\rho(\bvec{r})\!=\!\langle\!\langle\hat{\rho}\rangle\!\rangle$ is the average of the density operator $\hat{\rho}\!=\!\sum_{i=1}^N\delta(\rr-\rr_i)$.
Approximating now the inverse mobility matrix in Eq.~\eqref{eq_GammaU} as $\Gamma_{ij}^{-1}(\rr^N)\!\approx\!(\boldsymbol{1}-\ttau\nab_i\nab_j\mathcal{U})\delta_{ij}$
and integrating Eq.~\eqref{eq_ss1} over $N-1$ coordinates, we find 
the first member 
\begin{align}
\!\!-\rho(\rr)\langle\nab\beta \mathcal{U}\rangle
&=(1+\Da)\left(\nab\cdot\left(\boldsymbol{1}\rho(\rr)-\ttau\rho(\rr)\langle\Hess \mathcal{U}\rangle\right)\right)\,,
\label{eq_YBGss1}
\end{align}
of a YBG-like hierarchy~\cite{marconi2015,marconi2016} for the active system, where, explicitly
\begin{align}
\!\!\!\!\!\langle\mathfrak{D} \mathcal{U}\rangle=\mathfrak{D}\Vext(\rr)+
\int\upd\rr'\frac{\rho^{(2)}(\rr,\rr')}{\rho(\rr)}\mathfrak{D} u(\rr,\rr')\!\!\!\!\!\!\!
\label{eq_Gamma1av}
\end{align}
for any (nontrivial) differential operator $\mathfrak{D}_i$ acting on $\rr_i$.
In the derivation of \eqref{eq_YBGss1} it turns out that the off-diagonal components of the 
mobility tensor do not contribute at first order in $\tau$~\cite{marconi2016}.
Hence, we might as well have assumed the diagonal form $\Gamma_{ij}\approx\delta_{ij}\Gamma_{ii}$ at linear order in $\tau$ beforehand.

In order to connect to the EPA, 
we derive a YBG-like hierarchy from Eq.~\eqref{eq_ss2}.
Assuming the diagonal form $\Gamma_{ij}\approx\delta_{ij}\Gamma_{ii}$,
the integration over $N-1$ coordinates of the first equality is carried out i
n appendix \ref{app_Sumrules}.
Making use of the equilibrium version of the YBG hierarchy and
expanding the expression $\ln(\det\Gamma_{ii}(\rr^N))$ up to first order in $\tau$ the result is 
\begin{align}
0=-\boldsymbol{\mathcal{D}}_\text{I}^{-1}(\rr)\rho(\rr)\langle\nab\beta \mathcal{U}\rangle-\nab\rho(\rr)
+\ttau\rho(\rr)\langle\nab\cdot\Hess\mathcal{U}\rangle  +\frac{\rho(\rr)}{1+\Da}\,\ttau\int\upd\rr'\left(\Hess u(\rr,\rr')\right)\cdot\nab\frac{\rho^{(2)}(\rr,\rr')}{\rho(\rr)}\!\!
\label{eq_ss2int1}
\end{align}
introducing the averaged inverse diffusion tensor (compare Eq.~\eqref{eq_GammaU})
\begin{align}
\boldsymbol{\mathcal{D}}_\text{I}^{-1}(\rr):=
\frac{\left\langle\Gamma_{ii}\right\rangle}{1+\Da}=\frac{\boldsymbol{1}+\ttau\langle\Hess \mathcal{U}\rangle}{1+\Da}\,.
\label{eq_TeffINV}
\end{align}
Multiplying Eq.~\eqref{eq_ss2int1} with 
$\boldsymbol{\mathcal{D}}_\text{I}\approx(1+\Da)(\boldsymbol{1}-\ttau\langle\Hess \mathcal{U}\rangle)$ 
it is easy to verify in appendix \ref{app_Sumrules} that at first order in $\tau$ it
becomes equivalent to~\eqref{eq_YBGss1} up to a term proportional to the expression in the second line,
 which we consider as a higher-order contribution.

 In order to derive Eq.~\eqref{eq_ss2int1} in the Fox approach, an additional approximation is required,
 as the inverse of $\mathcal{D}^\FOX_{ij}$ from Eq.~\eqref{eq_GammaF} is not proportional to $\Gamma_{ij}$.
 We would thus need to redefine $\boldsymbol{\mathcal{D}}_\text{I}$ in Eq.~\eqref{eq_TeffINV}
 according to Eq.~\eqref{eq_GammaF} where $\left\langle\Gamma_{ii}\right\rangle^{-1}$
 takes the role of $\Gamma_{ij}^{-1}$.
Regarding in general the presented alternative derivation of Eq.~\eqref{eq_YBGss1},
its validity appears to be in question.
This is because to derive the intermediate result in Eq.~\eqref{eq_ss2int1}
 it is necessary to expand a logarithmic term,
the Taylor series of which only has a finite radius of convergence.
We further explicitly assumed the diagonal form of the mobility tensor
to avoid further terms that are not present in the original result.
 Employing in the next step the EPA will
shed more light on these issues.

\subsection{Low-activity approximation of effective potentials \label{sec_approximations}}

Having established a connection between \eqref{eq_ss1} and \eqref{eq_ss2} also at linear order in $\tau$,
we now turn to the case in which the second equality in \eqref{eq_ss2} does not hold.
This is when we assume $\bvec{F}_k^\text{eff}\!\approx\! -\nab_k\mathcal{U}^\text{eff}\!=\!-\nab_k\left(\Vext^\text{eff}(\rr_k)+
\sum_{l\neq k} u^\text{eff}(\rr_k,\rr_l)\right)$ 
along the lines of~\eqref{eq_Ftot} but within the EPA using the results derived in Sec.~\ref{sec_ueff1}. 
As detailed in appendix \ref{app_Sumrules2},
the obvious result is that all correlation functions between more than two particles vanish
in the approximate integrated version
\begin{align}
0=-\nab\rho(\rr)-\rho(\rr)\left\langle\nab\beta \mathcal{U}^\text{eff}\right\rangle
\label{eq_ss2int2}
\end{align}
of \eqref{eq_ss2}.
Ignoring the interparticle interactions the approximation involving only $\Vext^\text{eff}(\rr)$ becomes exact.
This situation is the same as discussed in Ref.~\onlinecite{marconi2015}.

Considering the interacting system, we multiply Eq.~\eqref{eq_ss2int2} with $\boldsymbol{\mathcal{D}}_\text{I}$
as done previously for~\eqref{eq_ss2int1}.
According to appendix \ref{app_Sumrules2}, we can only approximately reproduce Eq.~\eqref{eq_YBGss1} by doing so.
This reflects both the limitations of 
the EPA and an inconsistency between Eq.~\eqref{eq_ss1} and Eq.~\eqref{eq_ss2}
when being subject to the same type of approximation, as we discuss in the following. 
We observe that
(III.i) the coupling between external and internal interactions is ignored by Eq.~\eqref{eq_ss2int2}
(III.ii) spurious three-body correlations appear on the left-hand-side of Eq.~\eqref{eq_YBGss1}
(III.iii) the second term on the right-hand side of Eq.~\eqref{eq_YBGss1} is recovered but involves a seemingly unjustified expansion and
(III.iv) if we do not explicitly assume a diagonal diffusion tensor, the last term in Eq.~\eqref{eq_YBGss1} changes by a factor two.
As we are mainly interested in bulk systems, the first point is only briefly commented on in appendix \ref{app_Sumrules2}.

The term including the bare interaction force in Eq.~\eqref{eq_ss2} depends on the position of three bodies.
Hence, the pairwise approximation, which amounts to setting
\begin{align}
\sum_{l,j\neq k}\left(\nab_k\nab_ku(\rr_k,\rr_l)\right)\left(\nab_ku(\rr_k,\rr_j)\right)\;\longrightarrow\;
\sum_{j\neq k}\left(\nab_k\nab_ku(\rr_k,\rr_j)\right)\left(\nab_ku(\rr_k,\rr_j)\right)\,,
\end{align}
should not be too crude.
Moreover, we have discussed in Sec.~\ref{sec_ueff2}
that such a contribution to the effective force is usually purely repulsive and thus
plays only a minor role in characterizing a possible phase transition.
However, we show in appendix \ref{app_Sumrules2} that the definition~\eqref{eq_TeffINV}
of the averaged diffusion tensor $\boldsymbol{\mathcal{D}}_\text{I}$
is not fully compatible with the EPA, resulting in point (III.ii).
This is in contrast to the clean derivation of Eq.~\eqref{eq_YBGss1},
where Eq.~\eqref{eq_ss1} is recovered from Eq.~\eqref{eq_ss2}
 by multiplication with the many-body effective diffusion tensor $\mathcal{D}_{[N]}$
 before integrating over $N-1$ positions.

The last term $P_N\nab_k\ln|\det\mathcal{D}^{-1}_{kk}(\rr^N)|$ in \eqref{eq_ss2}, 
although considered here for a diagonal diffusion tensor,
 constitutes a full $N$-body quantity.
 Recall from the discussion in Sec.~\ref{sec_ueff} that
an approximation as a pairwise quantity might be quite poor and an expansion of the 
logarithm does not converge.
 However, we demonstrate in appendix \ref{app_Sumrules2} that successively employing the EPA and 
 expanding for small $\ttau$ according to 
 \begin{align}
\nab\ln\Bigg|\det&\left(\boldsymbol{1}+\ttau\Hess\sum_{l>1} u(\rr,\rr_l)\right)\Bigg|
\;\longrightarrow\;\nab\cdot\sum_{l>1}\ln|\det\left(\boldsymbol{1}+\ttau\Hess u(\rr,\rr_l)\right)|
\;\longrightarrow\;\ttau\sum_{l>1}\nab\boldsymbol{\Delta} u(\rr,\rr_l)+\mathcal{O}(\tau^2)
\label{eq_approximationschemeLOWtau}
\end{align}
 eventually results in full consistency with the respective term 
 in Eq.~\eqref{eq_YBGss1}, stated as point (III.iii).
 This suggests that the expansion to first order in $\tau$ "implies" making the EPA
when Eq.~\eqref{eq_ss2} is our starting point.
Despite the aforementioned crudity of this expansion, we argue that Eq.~\eqref{eq_YBGss1} is valid,
 as its clean derivation from Eq.~\eqref{eq_ss1} does not require dealing with a logarithmic term.
 The last step in Eq.~\eqref{eq_approximationschemeLOWtau} is required to recover Eq.~\eqref{eq_YBGss1}
without inducing undesired higher-order terms in $\tau$, 
 as, similar to point (III.ii), the integrated version is incompatible with the chosen $\boldsymbol{\mathcal{D}}_\text{I}$.
  However, we stress that this approximation
  should certainly be avoided when calculating the fluid structure.
  
Finally, we demonstrate in appendix \ref{app_Sumrules2}
 that the off-diagonal elements of the diffusion tensor entering in Eq.~\eqref{eq_ss2}
 contribute to Eq.~\eqref{eq_ss2int2}.
 Hence, the present approach would be even more inconsistent with Eq.~\eqref{eq_YBGss1}
 if we did not assume the diagonal form, as noted in point (III.iv).
 We also note that the same problem occurs for the according generalization of Eq.~\eqref{eq_ss2int1}. 
In principle we could define in this case an additional averaged diffusion 
tensor $\boldsymbol{\mathcal{D}}_\text{I,od}$, 
 corresponding to the off-diagonal elements,
 which could counteract this inconsistency.
Such a calculation would, however, not be useful when an effective pair potential is employed.

 \section{Conclusions \label{sec_conclusions}}

  In this paper we studied different ways to define an effective pair interaction potential between 
 active particles.
Our numerical investigation reveals that a two-particle system of ABPs and active OUPs
 exhibits a quite similar behavior.
These results serve as a benchmark to test the approximations involved in 
recent effective equilibrium approaches,
which have been reviewed and compared in detail.
For spatial dimensions higher than one
we introduced an empirical way to rid the theoretical result of possible divergences,
which also appears to yield the best agreement with the simulation data,
although the effective attraction is still significantly overestimated.
Regarding the quite accurate one-dimensional results
and the qualitative features of the effective potentials in three dimensions,
the Fox approximation is superior to the UCNA when the translational Brownian noise cannot be neglected.
In the absence of noise both approximation schemes admit the same steady-state solution.

 Further analysis is needed to better understand the role of the neglected many-body interactions
   in both the two-body simulations and the theory,
 which are thought to be imperative for a quantitative description of active systems~\cite{SpeckCRIT}.
The presented theoretical approach follows two major approximate steps to define the effective pair potential.
  First, we map the equation of motion~\eqref{eq_OUPs} onto a deterministic Fokker-Planck equation  
  (effective equilibrium picture)
 and then we define pair forces from the two-particle limit assuming a vanishing probability current. 
It could well be that the mapping in the first step breaks down 
 parts of the many-body nature of the interactions in the active system, 
 such that the effective attraction in the many-body system becomes accessible already on the 
 level of pair interactions.
  As a logical next step, it seems worthwhile to study the effective potential extracted 
from computer simulations of a many-particle system,
 in order to clarify in how far the strong attraction of the effective potential
 needs to be seen as the result of a fortuitous cancellation of errors.

 The low-activity limit in the effective equilibrium picture also results in pairwise forces.
 Under this assumption, we revealed some minor inconsistencies between the two equivalent steady-state conditions
  in Eqs.~\eqref{eq_ss1} and~\eqref{eq_ss2}, although the latter contains a logarithmic term.
  Relatedly, it was recognized in Ref.~\onlinecite{marconi2016} that different routes to define the active pressure 
only coincide at lowest order in the activity parameter $\tau$.
    We suspect that further differences will occur at higher orders in $\tau$ 
    and when employing further approximations, such as the EPA. 
We conclude that the route to follow should be carefully chosen for each problem,
together with the underlying approximations.

 The obvious purpose of both the low-activity approximation and the EPA
 is to allow for an analytically tractable theory.
It appears that the condition given by Eq.~\eqref{eq_ss2} supported by effective pair potentials is most
convenient for accessing structural properties~\cite{marconi2016mp,faragebrader2015,wittmannbrader2016},
whereas the low-activity expansion of Eq.~\eqref{eq_ss1} provides a direct way to define mechanical properties~\cite{marconi2016}.
Moreover, our analysis suggests that the thermodynamic results obtained from Eq.~\eqref{eq_ss2} can be rescaled in order to obtain
a workable definition of mechanical active pressure and surface tension.
Arguably, the most simplistic scaling factor would be the diffusivity $1+\Da$ of an ideal gas,
which can be absorbed into an effective temperature \cite{marconi2015,marconi2016,wittmannbrader2016}.
A more general approach will be detailed in the second paper of this series.

\section*{Acknowledgements}

R.\ Wittmann, A.\ Scacchi and J.\ M.\ Brader acknowledge funding provided by the Swiss National Science Foundation.
C.\ Maggi acknowledges support from the European Research Council under the European Union’s Seventh Framework programme
(FP7/2007-2013)/ERC Grant agreement
No. 307940.
\newline

\appendix

\section{Simulation details \label{app_SIM}}

 We performed Brownian Dynamics simulations of a system composed of two particles of unit diameter $d\!=\!1$ 
 interacting through a soft-repulsive potential or a Gaussian soft-core potential. 
 The potential is truncated at a distance of $r\!=\!2d$. In the simulations of active OUPs, evolving according to Eq.~\eqref{eq_OUPs}, 
 each particle is subjected to Gaussian thermal noise and a non-Gaussian (colored) noise.
 The latter yields two distinct contributions to the displacement of each particle: 
one drift term, proportional to the reorientation time $\tau$ and one Gaussian process, proportional to $\sqrt{\Da}/\tau$~\cite{sharma2016}. 
For a vanishing active diffusivity $\Da$, the drift term decays exponentially in time and is therefore irrelevant in the long-time limit.
 The integration time step is fixed to $\upd t\!=\!10^{-4}\tau_\text{B}$ where $\tau_\text{B}\!=\!d^2/D_\text{t}$ is the time scale of translational diffusion. 
 The total run time of the simulation is $10^6\tau_\text{B}$. For every $\upd t$, we calculate the distance between the two particles. 
 The pair-correlation function is obtained in a standard way from the distance distribution. 
  We have verified that, for the case of $\Da = 0$ and finite $\tau$, the obtained pair-correlation function is independent of $\tau$,
 although the short-time displacement is not.

We also performed Brownian Dynamics simulation of ABPs,
for which the colored-noise variable $\xchi_i(t)$ in Eq.~\eqref{eq_OUPs} is replaced with the vector $v_0\,\pvec_i(t)$ 
describing a constant velocity $v_0$ of the self-propulsion in the direction of the instantaneous orientation.
The equation $\dot{\pvec}_i(t)\!=\!\eeta_i(t)\times\pvec_i(t)$
for the time evolution for the orientation vector $\pvec_i(t)$ of each particle $i$ is evaluated as an Ito integral,
where $\eeta_i(t)$ is a white noise describing rotational diffusion. 
The integration time step is fixed to $\upd t=10^{-4}$ and the total run time is $10^4\tau_\text{B}$.

\section{Effective many-body force in the Fox approximation \label{app_DET}}

In this appendix we discuss the accuracy of Eq.~\eqref{eq_ss2} of the main text in the Fox approximation, i.e.,
choosing the effective diffusion tensor from Eq.~\eqref{eq_GammaF}.
The accurate definition of the effective force is
\begin{align}
\beta\bvec{F}_k^\text{eff}=\sum_i\mathcal{D}^{-1}_{ik}\cdot\beta\bvec{F}_i -\sum_{ij}\mathcal{D}^{-1}_{ik}\cdot\nab_j\cdot\mathcal{D}_{ji}
\label{eq_ss2detA}
\end{align}
since the conversion 
\begin{align}
 \sum_{ij}\mathcal{D}^{-1}_{ik}\cdot\nab_j\cdot\mathcal{D}_{ji}\approx\nab_k\ln|\det\mathcal{D}_{[N]}|
 \label{eq_ss2detAc}
\end{align}
is only correct in the following cases:
\begin{enumerate}
 \item for a system with no thermal noise.
 As stated in the main text, in this case the Fox and UCNA results are equivalent.
 Making use of the symmetry relation $\partial_{\gamma}\mathcal{D}^{-1}_{\alpha\beta}\!=\!\partial_{\beta}\mathcal{D}^{-1}_{\alpha\gamma}$ 
 (with Greek indices labeling components and particles)
 and Jacobi's formula
 the identity in Eq.~\eqref{eq_ss2detAc} can be explicitly verified~\cite{marconi2015}.
 \item for a passive system, since $\mathcal{D}_{ij}\!\equiv\!\boldsymbol{1}\delta_{ij}$.
 \item at leading order in the activity parameter $\tau$, where $\mathcal{D}^{-1}_{\alpha\beta}=\delta_{\alpha\beta}(1+\Da)-\ttau\Da\partial_\alpha\partial_\beta\,\mathcal{U}+\mathcal{O}(\ttau^2)$ and the same arguments as under point 1.\ can be used.
 \item for $N\!\leq\!2$ particles in an effectively one-dimensional symmetry, i.e., 
if there exists a coordinate frame in which the non-trivial contributions to
 $\mathcal{D}_{[N]}$ reduce to an at most a $2\times2$ tensor with identical diagonal elements.
This can be easily shown by an explicit calculation 
\begin{enumerate}
\item in $\di=1$ dimensions
\item for a planar interaction potential
\item in the Laplacian approximation~\eqref{eq_laplacian}
\end{enumerate}
\end{enumerate}

 As a simple counterexample to the cases listed above,
 we note that Eq.~\eqref{eq_ss2detAc} does not generally hold for $N\!=\!1$ and $\di\!=\!2$, since $\partial_x\Vext(x,y)\!\neq\!-\partial_y\Vext(x,y)$,
 whereas under point 4.\ we have $\partial_1u(x_1\!-\!x_2)\!=\!-\partial_2u(x_1\!-\!x_2)$.
 As for the approximate formulas~\eqref{eq_ueff} and~\eqref{eq_ueffDIAG}, we find that the difference between the
 effective forces, Eq.~\eqref{eq_ss2detA}, for $N\!=\!2$ particles with and without diagonal approximation
is only a factor 2 in front of each factor $\tau$.

\begin{figure}[t]{

\includegraphics[width=0.235\textwidth] {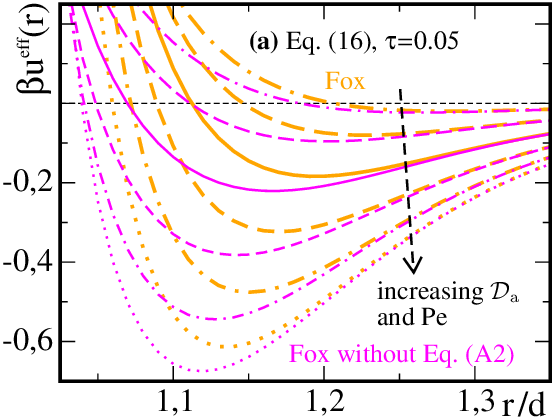} 
\includegraphics[width=0.235\textwidth] {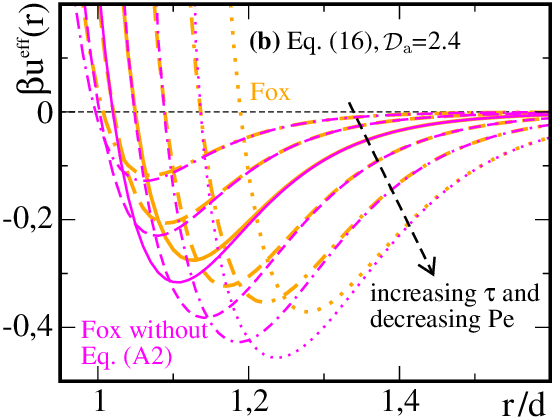}}

\caption{ Closeups of the Fox results for the full effective potentials from Eq.~\eqref{eq_ss2detAc} as in Fig.~\ref{fig_3d} \textbf{(a)} top right with additional data for $\Da\!=\!0.6$ and $\Da\!=\!0.3$ and \textbf{(b)} bottom right with additional data for $\tau\!=\!0.0125$ and $\tau\!=\!0.00625$. The thinner lines correspond to the results without making the approximation in Eq.~\eqref{eq_ss2detAc}.
\label{fig_3dA}}
\end{figure}

 Comparing the requirements for points 1.\ and 2.\ we can say that Eq.~\eqref{eq_ss2}
 is correct for both $\mathcal{D}_\text{a}\!\gg\!\mathcal{D}_\text{t}$ and $\mathcal{D}_\text{a}\!\ll\!\mathcal{D}_\text{t}$,
 indicating that it should be a good approximation over all ranges of the parameter $\mathcal{D}_\text{a}$.
 Moreover, the assumption of a small persistence time $\tau$ is required in the derivation of the effective equilibrium approach~\cite{faragebrader2015,maggi2015sr,marconi2015}.
 Considering point 3., this  means that the approximation in Eq.~\eqref{eq_ss2detAc} is consistent with the underlying theory.  
    Indeed, Figs.~\ref{fig_3dA}a and~\ref{fig_3dA}b show that the approximation best for either small or large $\Da$
   and small $\tau$, respectively.   
   In general, the difference is not significant compared to other approximations shown in Fig.~\ref{fig_3d} of the main text.
 
 If the validity criterion $E_n(\tau,r)\!>\!0$ for the Eigenvalues, given by Eq.~\eqref{eq_EVs},  of $\Gamma_{[2]}$ is violated, 
 neither side of Eq.~\eqref{eq_ss2detAc} results in physical effective potentials.
  Therefore, the most important benefit of the approximate form on the right-hand side is that it allows to employ 
  the inverse-$\tau$ approximation, as described in Sec.~\ref{sec_ueff2}:
  the Eigenvalues $E_n^\FOX(\tau,\Da,r)$ of $\mathcal{D}^\FOX_{[2]}$ can be written as 
  \begin{align}
E_n^\FOX(\tau,\Da,r)=1+\frac{\Da}{E_n(\tau,r)}
\end{align}
 so that we can substitute $E_n(\tau,r)$ according to Eq.~\eqref{eq_invtau} of the main text.
  The substitution of $E_n^\FOX$ or a more general manipulation of $\mathcal{D}^\FOX_{[2]}$ 
  is inconvenient since the effective potential would still diverge for $E_n\!=\!-\Da$ although the bare potential does not.
 Therefore, the third column of Fig.~\ref{fig_3d} contains the optimal implementation of the the inverse-$\tau$ approximation
 for the Fox approach.
 Also recall that, according to point 4.(c), the results in the Laplacian approximation are the same for both expressions in Eq.~\eqref{eq_ss2detAc}.

\section{Integration of the first equality in Eq.~\eqref{eq_ss2} \label{app_Sumrules}}

The derivation of Eq.~\eqref{eq_YBGss1} by integrating Eq.~\eqref{eq_ss1} over $N-1$ coordinates 
is quite similar to that of the YBG hierarchy in a passive system.
The first member 
\begin{align}
0=\nab\rho(\rr)+\rho(\rr)\nab\beta\Vext(\rr)+\int\upd\rr'\rho^{(2)}(\rr,\rr')\nab\beta u(\rr,\rr')=\rho(\rr)\nab\mu
\label{eq_YBGeq1}
\end{align}
is recovered from~\eqref{eq_YBGss1} when setting $\tau=\Da=0$.
The second equality reflects the interpretation of the term on the left-hand side
as the gradient of a chemical potential $\mu$, which is constant in equilibrium.
The second member reads
\begin{align}
0=\nab\rho^{(2)}(\rr,\rr'')+\rho^{(2)}(\rr,\rr'')\nab\left(\beta\Vext(\rr)+\beta u(\rr,\rr'')\right)+
\int\upd\rr'\rho^{(3)}(\rr,\rr',\rr'')\nab\beta u(\rr,\rr')=\rho^{(2)}(\rr,\rr'')\nab\mu\,.
\label{eq_YBGeq2}
\end{align}
and is related via the second equality to the first member.
With the help of these exact equilibrium sum rules
we will now derive Eq.~\eqref{eq_ss2int1} by integrating Eq.~\eqref{eq_ss2} over $N-1$ coordinates.
Our presentation follows closely the derivation of a dynamical density functional theory 
including a tensorial diffusivity~\cite{rexloewenDDFT},
whereas we only consider the steady-state condition.

We start by writing the first equality in~\eqref{eq_ss2} as 
\begin{align}
0=\sum_i\mathcal{D}^{-1}_{ik}\cdot\left(-\nab_iP_N+\beta\bvec{F}_i P_N\right)+\mathcal{D}^{-1}_{ik}\cdot\nab_iP_N-\nab_kP_N+P_N\nab_k\ln|\det\mathcal{D}^{-1}_{kk}|\,,
\label{eq_ss2A}
\end{align}
where we further used that the negative logarithm is the logarithm of the inverse argument and 
replaced $\det\mathcal{D}_{[N]}$ with $\det\mathcal{D}_{kk}$ by assuming the diagonal form. 
Integration of~\eqref{eq_ss2A} over $N-1$ coordinates yields (within UCNA)
\begin{align}
&\frac{1}{1+\Da}\left(-\nab\rho(\rr)-\rho(\rr)\nab\beta\Vext(\rr)-\int\upd\rr'\rho^{(2)}(\rr,\rr')\nab\beta u(\rr,\rr')\right.\cr
&\ \ \ \ \ \ \ \ \ \ \ \ 
-\ttau\Hess\Vext(\rr)\cdot\left(\nab\rho(\rr)+\rho(\rr)\nab\beta\Vext(\rr)+\int\upd\rr'\rho^{(2)}(\rr,\rr')\nab\beta u(\rr,\rr')\right)\cr
&\ \ \ \ \ \ \ \ \ \ \ \ 
-\ttau\int\upd\rr''\Hess u(\rr,\rr'')\cdot\left(\nab\rho^{(2)}(\rr,\rr'')+\rho^{(2)}(\rr,\rr'')\nab\left(\beta\Vext(\rr)
+\beta u(\rr,\rr'')\right)+\int\upd\rr'\rho^{(3)}(\rr,\rr',\rr'')\nab\beta u(\rr,\rr')\right)\cr
&\ \ \ \ \ \ \ \ \ \ \ \ \left.+\nab\rho(\rr)+\ttau\left(\Hess\Vext(\rr)\right)\cdot\nab\rho(\rr)
+\ttau\int\upd\rr'\left(\Hess u(\rr,\rr')\right)\cdot\nab\rho^{(2)}(\rr,\rr')\right)\cr
&-\nab\rho(\rr)+N\int\upd\rr_2\ldots\int\upd\rr_{N}P_N(\rr,\rr_2,\ldots,\rr_N) \nab\ln\left|\det\left(\boldsymbol{1}+\ttau\Hess\bigg(\Vext(\rr)+\sum_{l>1} u(\rr,\rr_l)\bigg)\right)\right|=0
\label{eq_ss2intA1}
\end{align}
As the Fox result~\eqref{eq_GammaF} for $\mathcal{D}^{-1}_{ik}$ is not a pairwise quantity,
a further approximation is required to obtain a similar hierarchy.
Now we eliminate the term in brackets within the third line containing the three-body correlation function 
with the help of \eqref{eq_YBGeq1} and \eqref{eq_YBGeq2} and
expand the logarithm up to linear order in $\tau$. The result is
\begin{align}
&-\frac{1}{1+\Da}\left(\boldsymbol{1}+\ttau\Hess\Vext(\rr)
+\ttau\int\upd\rr''\frac{\rho^{(2)}(\rr,\rr'')}{\rho(\rr)}\Hess u(\rr,\rr'')\right)\cdot\left(
\rho(\rr)\nab\beta\Vext(\rr)+\int\upd\rr'\rho^{(2)}(\rr,\rr')\nab\beta u(\rr,\rr')\right)\cr
&-\nab\rho(\rr)+\ttau\rho(\rr) \left(\nab\cdot\Hess\Vext(\rr)\right)
+\ttau\int\upd\rr'\rho^{(2)}(\rr,\rr')\nab\cdot\Hess u(\rr,\rr')\nonumber\\
& +\frac{1}{1+\Da}\,\ttau\int\upd\rr'\left(\Hess u(\rr,\rr')\right)\rho(\rr)\cdot\nab\frac{\rho^{(2)}(\rr,\rr')}{\rho(\rr)}=0
\label{eq_ss2intA3}
\end{align}
where we used the identity $\nab\boldsymbol{\Delta}=\nab\cdot\Hess$.
The term in the last line stems from replacing $\nab\rho^{(2)}(\rr,\rr'')$ in the third line of Eq.~\eqref{eq_ss2intA1}
with $\frac{\rho^{(2)}(\rr,\rr'')}{\rho(\rr)}\nab\rho(\rr)$, which does not cancel with the expression in the fourth line of Eq.~\eqref{eq_ss2intA1}.

For convenience we adopt the notational convention of the main text \eqref{eq_Gamma1av} and
identify the $\di\times\di$ matrix in the first line of Eq.~\eqref{eq_ss2intA3}
 as the inverse of an ensemble-averaged diffusion tensor
\begin{align}
\boldsymbol{\mathcal{D}}_\text{I}(\rr)&:=(1+\Da)\left(\boldsymbol{1}+\ttau\Hess\Vext(\rr)
+\ttau\int\upd\rr'\frac{\rho^{(2)}(\rr,\rr')}{\rho(\rr)}\Hess u(\rr,\rr')\right)^{-1}\cr&=
\left(\frac{\boldsymbol{1}+\ttau\langle\Hess \mathcal{U}\rangle}{1+\Da}\right)^{-1}=
(1+\Da)\left(\boldsymbol{1}-\ttau\langle\Hess \mathcal{U}\rangle\right)+\mathcal{O}(\tau^2)\,.
\label{eq_TeffINVapp}
\end{align}
Now we multiply Eq.~\eqref{eq_ss2intA3} with $\boldsymbol{\mathcal{D}}_\text{I}$ 
and drop all higher-order terms $\propto\tau^2$, which yields
\begin{align}
\rho(\rr)\langle\nab\beta \mathcal{U}\rangle&=(1+\Da)\left(-\nab\rho(\rr)+\ttau\left(\left(\nab\rho(\rr)\right)\cdot\langle\Hess \mathcal{U}\rangle
+\rho(\rr)\langle\nab\cdot\Hess \mathcal{U}\rangle\right)\right)
+{\ttau\int\upd\rr'\left(\Hess u(\rr,\rr')\right)\rho(\rr)\cdot\nab\frac{\rho^{(2)}(\rr,\rr')}{\rho(\rr)}}\cr
&=(1+\Da)\left(-\nab\rho(\rr)+\ttau\nab\cdot\left(\rho(\rr)\langle\Hess \mathcal{U}\rangle\right)\right)
-{\Da\ttau\int\upd\rr'\left(\Hess u(\rr,\rr')\right)\rho(\rr)\cdot\nab\frac{\rho^{(2)}(\rr,\rr')}{\rho(\rr)}}\,.
\label{eq_ss2intA4}
\end{align}
In the last step we made use of the identity
\begin{align}
\nab\cdot\left(\rho(\rr)\langle\Hess\beta \mathcal{U}\rangle\right)=\left(\nab\rho(\rr)\right)\cdot\langle\Hess\beta \mathcal{U}\rangle+\rho(\rr)\langle\nab\cdot\Hess\beta \mathcal{U}\rangle
+\rho(\rr)\int\upd\rr'\left(\Hess\beta u(\rr,\rr')\right)\cdot\nab\frac{\rho^{(2)}(\rr,\rr')}{\rho(\rr)}
\end{align}
to recover up to the last term the first member of the YGB-like hierarchy stated in Eq.~\eqref{eq_YBGss1} of the main text.
Taking into account the definition $\Da\propto\tau$ of the active diffusion coefficient,
we argue that the additional term is not relevant at linear order in $\tau$.
Alternatively, taking the mean-field approximation $\rho^{(2)}(\rr,\rr')\approx\rho(\rr)\rho(\rr')$,
this term will also vanish.
 We thus have rederived a result obtained in a much simpler way in Ref.~\onlinecite{marconi2015}.
The demonstrated equivalence of Eqs.~\eqref{eq_ss1} and~\eqref{eq_ss2} in
the low-activity limit is, however, not obvious and breaks down when higher-order terms in $\tau$ are included.

\section{Integration of the second equality in Eq.~\eqref{eq_ss2} \label{app_Sumrules2}}

Assuming pairwise interaction potentials, the integration of the second equality in Eq.~\eqref{eq_ss2} over $N-1$ coordinates
results in
\begin{align}
0&=\nab\rho(\rr)+\rho(\rr)\nab\beta\Vext^\text{eff}(\rr)+\int\upd\rr'\rho^{(2)}(\rr,\rr')\nab\beta u^\text{eff}(\rr,\rr')\cr
&=\nab\rho(\rr)+\rho(\rr)\left(\mathcal{D}^{-1}_{[1]}(\rr)\nab\beta \Vext(r)-\nab\ln|\det\mathcal{D}_{[1]}(\rr)|\right)\cr
&\ \ \ +\left[\int\upd\rr'\rho^{(2)}(\rr,\rr')\left(\mathcal{D}^{-1}_{11}(\rr,\rr')\cdot\nab\beta u(\rr,\rr')
-\nab\ln|\det\mathcal{D}_{11}(\rr,\rr')|\right)\right]_{\Vext=0}
\label{eq_ss2intB1}
\end{align}
where in the second step we have inserted the effective external~\eqref{eq_Veff}
and pair potential~\eqref{eq_ueffDIAG}.
To be consistent with appendix~\ref{app_Sumrules} we used the diagonal form of the latter.
In the absence of interparticle interactions, $u(r)=0$, 
it is easy to verify that both Eqs.~\eqref{eq_ss2intB1} and~\eqref{eq_ss2intA1} simplify to the same equality
\begin{align}
\nab\rho(\rr)+\rho(\rr)\left(\frac{\left(\nab\beta\Vext(\rr)\right)\cdot\left(\boldsymbol{1}
+\ttau\Hess\Vext(\rr)\right)}{1+\Da}
-\nab\ln|\det\left(\boldsymbol{1}+\ttau\Hess\Vext(\rr)\right)|\right)=0\,,
\label{eq_ss2intB2V}
\end{align}
which is a trivial consequence of the fact that the many-body potential is the sum of
single-particle contributions: the friction tensor \eqref{eq_Gamma0} is diagonal and all equations decouple.
Comparing the result~\eqref{eq_ss2intB1} in the interacting case to Eq.~\eqref{eq_ss2intA3},
we notice that the EPA ignores the cross terms proportional to
$\ttau\Hess\Vext(\rr)\cdot\int\upd\rr'\rho^{(2)}(\rr,\rr')\nab\beta u(\rr,\rr')$ and 
$\ttau\nab\beta\Vext(\rr)\cdot\int\upd\rr'\rho^{(2)}(\rr,\rr')\Hess u(\rr,\rr')$,
coupling the external and internal interactions on the level of pair correlations. 
It is, however, possible to capture these terms
within a generalized effective external two-body field in the spirit of Ref.~\onlinecite{marconi2016mp}.
This amendable difference aside, we now discuss the bulk system.

Setting $\Vext(\rr)=0$ in Eq.~\eqref{eq_ss2intB1} becomes
\begin{align}
&\nab\rho(\rr)+\int\upd\rr'\rho^{(2)}(\rr,\rr')\left(\frac{\left(\nab\beta u(\rr,\rr')\right)\cdot
\left(\boldsymbol{1}+\ttau\Hess u(\rr,\rr')\right)}{1+\Da}
-\nab\ln|\det\left(\boldsymbol{1}+\ttau\Hess u(\rr,\rr')\right)|\right)=0
\label{eq_ss2intB3U}
\end{align}
This result amounts to setting
\begin{align}
&\frac{\ttau}{1+\Da}\int\upd\rr''\left(\nab\beta u(\rr,\rr'')\right)\cdot
\int\upd\rr'\rho^{(3)}(\rr,\rr',\rr'')\Hess u(\rr,\rr')\rightarrow0\,,\label{eq_ss2intB3Ua}\\
&N\int\upd\rr_2\ldots\upd\rr_{N}P_N(\rr^N) \nab\ln\left|\det\left(\boldsymbol{1}+\ttau\Hess\sum_{l>1} u(\rr,\rr_l)\right)\right|
\rightarrow\int\upd\rr'\rho^{(2)}(\rr,\rr')\nab\ln|\det\left(\boldsymbol{1}+\ttau\Hess u(\rr,\rr')\right)|
\label{eq_ss2intB3Ub}
\end{align}
in~\eqref{eq_ss2intA1},
which is a logical consequence of the higher-order correlations being ignored.
Restricting ourselves to the leading order in $\tau$ both sides of \eqref{eq_ss2intB3Ub}
reduce to the equivalent form $\ttau\int\upd\rr'\rho^{(2)}(\rr,\rr')\nab\cdot\Hess u(\rr,\rr')$.
In contrast, making use of \eqref{eq_YBGeq2}, the approximation \eqref{eq_ss2intB3Ua} to the bare force term is equivalent to setting
\begin{align}
&\frac{\ttau}{1+\Da}\int\upd\rr''\frac{\rho^{(2)}(\rr,\rr'')}{\rho(\rr)}\Hess u(\rr,\rr'')\cdot\int\upd\rr'\rho^{(2)}(\rr,\rr')\nab\beta u(\rr,\rr')
+{\frac{\ttau}{1+\Da}\int\upd\rr'\left(\Hess u(\rr,\rr')\right)\rho(\rr)\cdot\nab\frac{\rho^{(2)}(\rr,\rr')}{\rho(\rr)}}\nonumber\\
&\rightarrow\frac{\ttau}{1+\Da}\int\upd\rr'\rho^{(2)}(\rr,\rr')(\nab\beta u(\rr,\rr'))\cdot(\Hess u(\rr,\rr'))
\label{eq_ss2intB3Uax}
\end{align}
in \eqref{eq_ss2intA3}. 
Thus the factorization in the first line of \eqref{eq_ss2intA3} is not possible,
such that, upon multiplying with $\boldsymbol{\mathcal{D}}_\text{I}$ defined in Eq.~\eqref{eq_TeffINVapp} there remains an additional term 
\begin{align}
\ttau\int\upd\rr'\rho^{(2)}(\rr,\rr')(\nab\beta u(\rr,\rr'))\cdot\left(\Hess u(\rr,\rr')-
\int\upd\rr''\frac{\rho^{(2)}(\rr,\rr'')}{\rho(\rr)}\Hess u(\rr,\rr'')\right)
\label{eq_ss2intB3UaxC}
\end{align}
proportional to $\tau$ on the left-hand-side of Eq.~\eqref{eq_ss2intA4}. 
This means that the EPA
introduces a three-body term to the YBG-like hierarchy~\eqref{eq_YBGss1}.
The reason for this discrepancy is that the effective diffusion tensor $\boldsymbol{\mathcal{D}}_\text{I}$ is defined independently of
the approximation made in Eq.~\eqref{eq_ss2intB3Ua}.

 Finally, we note that if we employ in Eq.~\eqref{eq_ss2intB1} the effective pair potential~\eqref{eq_ueff}
  that does not correspond to a diagonal diffusion tensor,
 we will have to modify Eq.~\eqref{eq_ss2intB3U} by setting $\tau\rightarrow2\tau$.
This reflects in general the inconsistency between the low-activity expansions 
of the two versions of the steady-state condition given by Eqs.~\eqref{eq_ss1} and~\eqref{eq_ss2},
which is not a consequence of approximating the effective force in the second form using pair potentials.
In particular, we would also have to substitute $\tau\rightarrow2\tau$ in Eq.~\eqref{eq_ss2intA4},
as both sides in Eq.~\eqref{eq_ss2intB3Ub} are equivalent at linear order in $\tau$.
However, in Eq.~\eqref{eq_ss2intB3UaxC} only the first term should then be multiplied by the factor two,
as the second term arises from $\boldsymbol{\mathcal{D}}_\text{I}$ and not from the effective potential.

\end{document}